\documentclass[useAMS,usenatbib]{mn2e}
\usepackage{epsfig}
\usepackage{amsfonts}
\usepackage[usenames,dvipsnames]{xcolor}
\usepackage{url}
\usepackage{subfigure}
\usepackage{times,graphicx,amsmath,amsfonts,amssymb,aas_macros,epstopdf,hyperref}
\usepackage[normalem]{ulem}
\usepackage[T1]{fontenc}
\usepackage{multirow}

\def\ie{{\frenchspacing\it i.e.}}
\def\eg{{\frenchspacing\it e.g.}}

\def\be{\begin{equation}}
\def\ee{\end{equation}}
\def\ba{\begin{eqnarray}}
\def\ea{\end{eqnarray}}

\newcommand{\fcb}{f_{\rm cb}}
\newcommand{\fnu}{f_{\nu}}

\newcommand{\hompc}{\,h\,{\rm Mpc}^{-1}}

\def\LaTeX{L\kern-.36em\raise.3ex\hbox{a}\kern-.15em
    T\kern-.1667em\lower.7ex\hbox{E}\kern-.125emX}

\begin{document}

\voffset-1.25cm
\title[Weighing the neutrino mass using the CMASS power spectrum]{The clustering of galaxies in the SDSS-III Baryon Oscillation Spectroscopic Survey: weighing the neutrino mass using the galaxy power spectrum of the CMASS sample}
\author[Zhao, Saito, Percival et al.]{
\parbox{\textwidth}{
Gong-Bo Zhao$^{1,2}$\thanks{Email: Gong-Bo.Zhao@port.ac.uk}, Shun Saito$^{3,4}$, Will J. Percival$^{1}$, Ashley J. Ross$^{1}$, Francesco Montesano$^{5}$, Matteo Viel$^{6,7}$, Donald P. Schneider$^{8,9}$, Marc Manera$^{1}$,  Jordi Miralda-Escud\'e$^{10,11}$, Nathalie Palanque-Delabrouille$^{12}$, Nicholas P. Ross$^4$, Lado Samushia$^{1,13}$, Ariel G. S{\'a}nchez$^5$, Molly E.~C.~Swanson$^{14}$, Daniel Thomas$^{1}$, Rita Tojeiro$^{1}$, Christophe Y\`eche$^{12}$, Donald G. York$^{15}$
}
\vspace*{6pt} \\
$^1$ Institute of Cosmology \& Gravitation, University of Portsmouth, Dennis Sciama Building, Portsmouth, PO1 3FX, UK\\
$^2$ National Astronomy Observatories, Chinese Academy of Science, Beijing, 100012, P.R.China\\
$^3$ Department of Astronomy, 601 Campbell Hall, University of California at Berkeley, Berkeley, CA 94720, U.S.A.\\
$^4$ Lawrence Berkeley National Laboratory, 1 Cyclotron Road, Berkeley, CA 92420, U.S.A.\\
$^5$ Max-Planck-Institut f\"ur extraterrestrische Physik, Postfach 1312, Giessenbachstr., 85748 Garching, Germany\\
$^6$ INAF-Osservatorio Astronomico di Trieste, Via G.B. Tiepolo 11, I-34131 Trieste, Italy\\
$^7$ INFN/National Institute for Nuclear Physics, Via Valerio 2, I-34127 Trieste, Italy\\
$^8$ Department of Astronomy and Astrophysics, The Pennsylvania State University, University Park, PA 16802, U.S.A.\\
$^9$ Institute for Gravitation and the Cosmos, The Pennsylvania State University, University Park, PA 16802, U.S.A.\\
$^{10}$ Instituci\'o Catalana de Recerca i Estudis Avan\c{c}ats, Barcelona, Catalonia\\
$^{11}$ Institut de Ci\`encies del Cosmos, Universitat de Barcelona/IEEC, Barcelona, Catalonia\\
$^{12}$ CEA, Centre de Saclay, IRFU, 91191 Gif-sur-Yvette, France\\
$^{13}$ National Abastumani Astrophysical Observatory, Ilia State University, 2A Kazbegi Ave., GE-1060 Tbilisi, Georgia\\
$^{14}$ Harvard-Smithsonian Center for Astrophysics, 60 Garden St., MS 20, Cambridge, MA 02138, U.S.A.\\
$^{15}$ Department of Astronomy and Astrophysics and the Enrico Fermi Institute, University of Chicago, 5640 South Ellis Avenue, Chicago, IL 60637, U.S.A.}
\date{\today} 
\pagerange{\pageref{firstpage}--\pageref{lastpage}}

\label{firstpage}

\maketitle

\begin{abstract} 

  We measure the sum of the neutrino particle masses using the three-dimensional galaxy power spectrum of the SDSS-III Baryon Oscillation Spectroscopic Survey (BOSS) Data Release 9 (DR9) CMASS galaxy sample. Combined with the cosmic microwave background (CMB), supernova (SN) and additional baryonic acoustic oscillation (BAO) data, we find upper 95 percent confidence limits of the neutrino mass $\Sigma m_{\nu}<0.340$\,eV within a flat $\Lambda$CDM background, and $\Sigma m_{\nu}<0.821$\,eV, assuming a more general background cosmological model. The number of neutrino species is measured to be $N_{\rm eff}=4.308\pm0.794$ and $N_{\rm eff}=4.032^{+0.870}_{-0.894}$ for these two cases respectively. We study and quantify the effect of several factors on the neutrino measurements, including the galaxy power spectrum bias model, the effect of redshift-space distortion, the cutoff scale of the power spectrum, and the choice of additional data. The impact of neutrinos with unknown masses on other cosmological parameter measurements is investigated. The fractional matter density and the Hubble parameter are measured to be $\Omega_{\rm M}=0.2796\pm0.0097,~H_0=69.72^{+0.90}_{-0.91}$ km/s/Mpc (flat $\Lambda$CDM) and $\Omega_{\rm M}=0.2798^{+0.0132}_{-0.0136},~H_0=73.78^{+3.16}_{-3.17}$ km/s/Mpc (more general background model). Based on a Chevallier-Polarski-Linder (CPL) parametrisation of the equation-of-state $w$ of dark energy, we find that $w=-1$ is consistent with observations, even allowing for neutrinos. Similarly, the curvature $\Omega_K$ and the running of the spectral index $\alpha_s$ are both consistent with zero. The tensor-to-scalar ratio is constrained down to $r<0.198$ (95 percent CL, flat $\Lambda$CDM) and $r<0.440$ (95 percent CL, more general background model).

\end{abstract}



\section{Introduction}
\label{sec:intro}

The determination of the neutrino mass is one of the most important tasks in modern science. High-energy experiments probe the mass differences between neutrino species through neutrino oscillations. Latest measurements give squared mass differences for the solar neutrino and the atmospheric neutrinos of $\Delta m_{12}^2=7.59^{+0.20}_{-0.18}\times10^{-5}$\,eV$^2$ and $\Delta m_{23}^2=2.35^{+0.12}_{-0.09}\times10^{-3}$\,eV$^2$, respectively \citep{Schwetz11,Fogli11}. 

While there is no way to measure the absolute neutrino mass in these oscillation experiments this can be measured using cosmological observations. The redshift of the matter-radiation equality depends on the summed mass of neutrino particles, leaving an imprint on the cosmic microwave background (CMB) angular power spectrum, and the matter transfer function. Massive neutrinos also affect the cosmological growth factor, inhibiting growth as their velocity dispersion slows down their mean flow into structures. Reviews of the effects of neutrinos in cosmology are given by \citet{Dolgov02} and \citet{Lesgourgues06}.

Given the number of recent galaxy surveys, and different options for modelling non-linear effects in the measured galaxy power spectrum, it is not surprising that there have been a significant number of previous efforts to measure the neutrino mass using cosmological probes, (\eg, \citealt{Elgaroy02, Lewis:2002ah, Allen03, Hannestad03, Spergel03, Barger04, Crotty04, Hannestad04, Tegmark04, Elgaroy05, Hannestad05, Seljak05, Goobar06, Spergel07, Xia07, Komatsu09, Tereno09, Reid10a, Reid10b, Saito:2010, Swanson2010, Thomas10, Benson11, Komatsu11, Riemer11, dePutter:2012sh, Xia12, Sanchez:2012sg}). In this work, we use the best-ever measurement of the galaxy power spectrum, based on the SDSS-III Baryon Oscillation Spectroscopic Survey (BOSS) combined with other datasets of CMB, SNe and BAO to measure the neutrino mass, the effective number of species of the neutrinos, as well as other degenerate cosmological parameters taking into account the neutrino effects.

We shall briefly introduce the BOSS sample in the next section, and describe the method adopted to model the galaxy power spectrum in Section 3. Before we show our results in Section 5, we detail the datasets included in addition to the BOSS data. We conclude and review the primary results of this work in the final section. 

\section{CMASS DR9 data and measurement}

\label{sec:data}
The SDSS-III \citep{eisenstein11} Baryon Oscillation Spectroscopic Survey \citep{Dawson12} obtains spectra for two primary galaxy samples selected from SDSS imaging data. In combination, the SDSS-I, SDSS-II, and SDSS-III imaging surveys obtained wide-field CCD photometry (\citealt{Gunn98,Gunn06}) in five passbands ($u,g,r,i,z$; e.g., \citealt{Fukugita96}), amassing a total footprint of 14,555 deg$^2$, internally calibrated using the `uber-calibration' process described in \cite{Pad08}, and with a 50 percent completeness limit of point sources at $r = 22.5$ (\citealt{DR8}). From this imaging data, BOSS has targeted 1.5 million massive galaxies split between LOWZ and CMASS samples over an area of 10,000 deg$^2$ (Padmanabhan et al. in prep.). BOSS observations began in fall 2009, and the last spectra of targeted galaxies will be acquired in 2014. The R = 1300-3000 BOSS spectrographs \citep{2012arXiv1208.2233S} are fed by 1000 2$^{\prime\prime}$ aperture optical fibres, allowing 1000 objects to be observed in a single pointing. Each galaxy observation is performed in a series of 15-minute exposures and integrated until a fiducial minimum signal-to-noise ratio, chosen to ensure a high redshift success rate, is reached (see \citealt{Dawson12} for further details). The determination of spectroscopic redshifts from these data is described in \citet{Bolton12}.

We use the SDSS-III Data Release 9 (DR9; \citealt{2012arXiv1207.7137S}) CMASS sample of galaxies, as analysed by \cite{alph, ReidDR9,Samushia12,Sanchez:2012sg,Nuza12,Ross12a,Ross12b}. It contains $264\,283$ massive galaxies covering 3275\,deg$^2$ with redshifts $0.43<z<0.7$ (the effective redshift $z_{\rm eff}=0.57$. The sample is roughly split into two angular regions, with 2635\,deg$^2$ in the Northern Galactic Cap and 709\,deg$^2$ in the Southern Galactic Cap. 

We measure the galaxy power spectrum $P_{\rm meas}(k)$, using the standard Fourier technique of \cite{FKP}, as described by \cite{ReidDR7}. In particular, we calculate the spherically-averaged power in $k$ bands of width $\Delta k = 0.004\hompc$ using a 2048$^3$ grid. We use the same weighting scheme as in \cite{alph}, \cite{Ross12a} and \cite{Ross12b}, which includes both weights to correct for the systematic relationship between target galaxy density and stellar density and also `FKP' weights, using the prescription of \cite{FKP}, which amounts to a redshift-dependent weighting in our application. 
The process of calculating weights is described in detail in \cite{Ross12a}. The galaxy power spectrum was used in \citet{alph} to extract and fit the Baryon Acoustic Oscillations. 

To determine expected errors on the power spectrum, we use the 600 mock DR9 CMASS catalogues generated by \cite{Manera12} to generate a covariance matrix $C[k_i][k_j]$ quantifying the covariance of the power between the $i$th and $j$th $k$ bins. The correlation between bins can be simply quantified using the correlation matrix Corr$[k_i][k_j]$, 
\begin{equation} \label{eq:corr}
  {\rm Corr}[k_i][k_j]\equiv\frac{C[k_i][k_j]}{\sqrt{C[k_i][k_i]C[k_j][k_j]}}. 
\end{equation} 
\cite{Manera12} used the initial conditions of a flat cosmology defined by $\Omega_{\rm M}=0.274, \Omega_{b}h^2=0.0224, h=0.70, n_s=0.95$, and $\sigma_8=0.8$ (matching the fiducial cosmologies assumed in \citealt{White11BEDR} and \citealt{alph}), and generated dark matter halo fields at redshift 0.55. These were produced using a 2nd-order Lagrangian Perturbation Theory (2LPT) approach inspired by the Perturbation Theory Halos (PTHalos; \citealt{PTHalos}). Galaxies were placed in halos using the halo occupation distribution determined from CMASS $\xi_0$ measurements and the parameterization of \cite{ZCZ}. The DR9 angular footprint was then applied and galaxies were sampled along the radial direction such that the mean $n(z)$ matched the CMASS $n(z)$, thereby providing 600 catalogues simulating the observed DR9 CMASS sample. See \cite{Manera12} for further details.

\begin{figure}
  \begin{center}
  \includegraphics[scale=0.27]{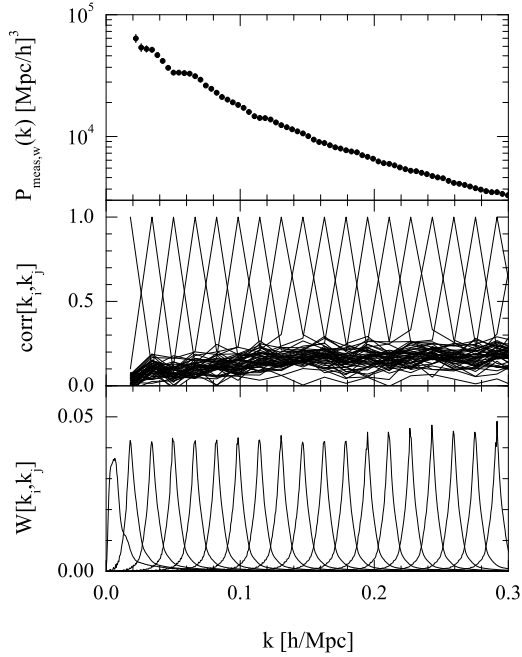}
  \end{center}
  \caption{The power spectrum measured from the CMASS sample. The correlation coefficients obtained from the mock catalogues and the expected window functions are shown in the middle and bottom row. See text for more details.} \label{fig:data}
\end{figure}

\citet{Ross12a} and \citet{Ross12b} consider remaining systematic uncertainties in the power spectrum measurements, caused by observational effects \footnote{The systematic uncertainties only affect the power spectrum measurements on large scales ($k<0.02 \hompc$).}. They suggest that the most conservative treatment is to allow for a free parameter in the measured data such that
\begin{equation}
P_{\rm meas}(k) = P_{\rm meas,w}(k)-S[P_{\rm meas,nw}(k)-P_{\rm meas,w}(k)],
\label{eq:pmeasu}
\end{equation}
where $P_{\rm meas,w}(k)$ is the power spectrum measured after applying the weights for stellar density (the dominant known source of systematic deviations in target density) and $P_{\rm meas,nw}(k)$ is the measurement without these weights. The parameter $S$ represents an additional nuisance parameter that we marginalise over for all of the constraints we present. Note that fixing $S=0$ represents the case where no systematic uncertainty in the application of the weights for stellar density is taken into account when measuring the power spectrum using these weights. Following \citet{Ross12b}, we put a Gaussian prior on $S$ around zero with a standard deviation $0.1$,  which is motivated by the tests using mocks in \citet{Ross12a}.

In the resulting measurements, data at different scales will be correlated as a result of the survey geometry. We must account for this effect when comparing theoretical power spectra, $P_g(k)$ (calculated as described in Section 3), to $P_{\rm meas}(k)$. We do so using the spherically-averaged power in the window, which we denote $P_{\rm win}(k)$: the measured power spectrum $P_{\rm meas}(k)$ is a convolution of the true underlying power spectrum with $P_{\rm win}(k)$, normalised such that the convolved product is zero at $k=0$. For ease we convert the convolution into a matrix multiplication based on a ``window matrix'' $W[k_i][k_j]$ such that
\begin{equation}
P_{\rm conv}(k_i) = \sum_j W[k_i][k_j]P{_g}(k_j) - P_oP_{\rm win}(k)
\label{eq:pkm}
\end{equation}
where
\begin{equation}
P_o = \sum_jW[0][k_j]P_{g}(k_j)/P_{\rm win}(0).
\label{eq:p0}
\end{equation}
The 2nd term in Eq. (\ref{eq:pkm}) is necessary because the galaxy density is estimated from the data itself, forcing $P_{\rm meas}(k=0)=0$. We calculate $P_{\rm win}(k)$ and $P_g(k)$ in bins $\Delta k = 0.0005\hompc$, yielding sufficient resolution to output $P_{\rm conv}(k)$ in (the measured) bin width $\Delta k = 0.004\hompc$. \cite{Ross12a} present a number of tests of this procedure based on analysis of the mock samples. The $P_{\rm conv}(k)$ we determine using this approach are compared directly to $P_{\rm meas}(k)$.

We assume a flat $\Lambda$CDM cosmology with $\Omega_{\rm M} = 0.285,~\Omega_b = 0.0459,~h = 0.70,~n_s = 0.96,~\sigma_8 = 0.8$ (approximately the best-fit model found by \citealt{Sanchez:2012sg}) when calculating the power spectrum, and we project the scales (using $D_V(z=0.57)$ as the distance) to the cosmology to be tested. The quantity $D_V$ is the average distance to the galaxies-pairs, defined as, \be\label{eq:DV} D_V(z)=\left[(1+z)^2 D_A^2(z) \frac{cz}{H(z)}\right]^{1/3} \ee where $D_A(z)$ and $H(z)$ are the physical angular diameter distance and 
the Hubble expansion rate at redshift $z$ respectively. 

The measured power spectra $P_{\rm meas,w}(k)$, the correlation matrix Corr$[k_i][k_j]$ (defined in Eq. \ref{eq:corr}) and the window matrix $W[k_i][k_j]$ are shown in Fig.~\ref{fig:data} (for clarity, only every 4th $k$-bin are displayed in the middle and bottom panels). The effect of the survey window is clear, as the middle panels show that there is significant correlation between $k$-bins. These correlations are larger in the Southern region, as is the uncertainty, due to the fact that it accounts for only 22\% of the area observed in the total combined sample.

\begin{table*}
\begin{center}
\begin{tabular}{cccc}

\hline\hline
Parameter                                          &  Meaning                                           &  Prior (flat)    &   Value for the vanilla model \\ \hline
$\omega_{b}\equiv\Omega_{b}h^2$  & The physical baryon energy density & $[0.005,0.1]$        & varied\\
$\omega_{c}\equiv\Omega_{c}h^2$  & The physical dark matter energy density & $[0.01,0.99]$        & varied\\
$\Theta_{s}$  & $100\times$the ratio of the sound horizon to the angular diameter distance at decoupling & $[0.5,10]$       & varied\\
$\tau$  & The optical depth & $[0.01,0.2]$        & varied\\
$n_s$  & The spectral index of the primordial power spectrum & $[0.5,1.5]$        & varied\\
log$[10^{10}A_s]$  & The amplitude of the primordial power spectrum & $[2.7,4]$        & varied\\
$A_{\rm SZ}$  & The amplitude of SZ power spectrum when using CMB & $[0,2]$        & varied\\
$\alpha_s$  & The running of the primordial power spectrum & $[-0.1,0.1]$        & 0\\
$\Sigma m_{\nu}$  & The sum of the neutrino masses in the unit of eV & $[0,2]$        & 0\\
$N_{\rm eff}$  & The number of the neutrino species  & $[1.5,10]$        & 3.046\\
$w_0$  & The $w_0$ parameter in the CPL parametrisation  & $[-3,3]$        & $-1$\\
$w_a$  & The $w_a$ parameter in the CPL parametrisation  & $[-3,3]$        & $0$\\
$\Omega_K$  & The contribution of the curvature to the energy density  & $[-0.1,0.1]$        & $0$\\
$r$  & The tensor to scalar ratio  & $[0,2]$        & $0$\\ \hline
$\alpha$  & The nuisance parameter for SN defined in Eq.~(\ref{eq:sn})  & $[0.6,2.6]$        & varied\\
$\beta$  & The nuisance parameter for SN defined in Eq.~(\ref{eq:sn})  & $[0.9,4.6]$        & varied\\
$b_1$  & The nuisance parameter for $P(k)$ when using Eq.~(\ref{eq:nonlinearPkhalo})  & $[1,3]$        & varied\\
$b_2$  & The nuisance parameter for $P(k)$ when using Eq.~(\ref{eq:nonlinearPkhalo})  & $[-4,4]$        & varied\\
$N$  & The nuisance parameter for $P(k)$ when using Eq.~(\ref{eq:nonlinearPkhalo})  & $[0,5000]$        & varied\\
$b_{\rm HF}$  & The nuisance parameter for $P(k)$ when using Eq.~(\ref{eq:bHF})  & $[0.1,10]$        & varied\\
$P_{\rm HF}$  & The nuisance parameter for $P(k)$ when using Eq.~(\ref{eq:bHF})  & $[0,5000]$        & varied\\
$b_Q$  & The nuisance parameter for $P(k)$ when using Eq.~(\ref{eq:bQ})  & $[0.1,10]$        & varied\\
$Q$  & The nuisance parameter for $P(k)$ when using Eq.~(\ref{eq:bQ})  & $[0.1,50]$        & varied\\
$S$  & The nuisance parameter for $P(k)$ measurement systematics, defined in Eq.~(\ref{eq:pmeasu})  & $[-1,1]$        & varied\\

\hline\hline

\end{tabular}
\end{center}
\caption{The parameters used in our analysis and their physical meaning, ranges and values for the vanilla $\Lambda$CDM model. }
\label{tab:param}
\end{table*}%

\section{Modelling the galaxy power spectrum} 

As massive neutrinos change the underlying shape of the matter power spectrum, resulting constraints on non-linear scales depend on our ability to model similar changes in shape induced by the physics of galaxy formation - known as galaxy bias. Various models and parameterisations have previously been introduced to convert from the linear matter power spectrum to the galaxy power spectrum. \citet{Swanson2010} tried twelve different models and compared the resultant neutrino mass measurement using SDSS-II LRG sample. Models are based on two different approaches: either using higher-than-linear order perturbation theory \eg, \citealt{Saito:2008lr,Saito:2009fk}), or using fitting formulae calibrated from $N$-body simulations (\eg, {\sc HALOFIT}; \citealt{Smith2003,Bird2012}). In this work, we shall compare results from both approaches. 

\subsection{Perturbation theory}\label{sec:PT}

To extend the power spectrum to mildly nonlinear scales, a straightforward approach is to include higher-order corrections to the linear standard perturbation theory (SPT). For the system consisting of baryons and dark matter only, the higher-order corrections (dubbed `one-loop' corrections) have been extensively studied (\eg, \citealt{Juszkiewicz81,Makino92,jain94,Scoccimarro96,Heavens98}). 
At the one-loop level of the SPT, the corrected matter power spectrum is
\begin{equation}\label{eq:1loop} 
  P^{\rm 1loop}_{\rm cb}(k)\equiv{P^{\rm L}_{\rm cb}}(k)+{P^{(22)}_{\rm cb}}(k)+{P^{(13)}_{\rm cb}}(k),
\end{equation} 
where the subscript `cb' denotes `CDM plus baryons', the superscripts `(22)' and `(13)' illustrate the perturbative corrections to the power spectrum at next-to-leading order, and the superscript `L' stands for the linear matter power spectrum. With the presence of massive neutrinos, \citet{Saito:2008lr,Saito:2009fk} generalised Eq.~(\ref{eq:1loop}) to \begin{equation}\label{eq:mix} 
  P^{\rm 1loop}_{\rm cb\nu}(k)=\fcb^{2}P^{\rm 1loop}_{\rm cb}(k)+2\fcb\fnu P^{\rm L}_{\rm cb\nu}(k) 
    + \fnu^{2} P^{\rm L}_{\nu}(k),
\end{equation} 
where $P^{\rm L}_{\rm cb\nu}(k)$ and $P^{\rm L}_{\nu}(k)$ represent the linear power spectrum for total matter (CDM plus baryons plus massive neutrinos) and for massive neutrinos only, respectively, and the coefficient $f_i$ denotes the mass
fraction of each species relative to the present-day energy density of
total matter, $\Omega_{\rm m}$, \ie, 
\begin{equation} 
  f_{\nu}\equiv\frac{\Omega_{\nu}}{\Omega_{\rm m}} 
    =\frac{\sum m_{\nu}}{\Omega_{\rm m} h^2 \times 94.1{\rm eV}},~f_{\rm cb}=1-f_\nu. 
\end{equation} 
A crucial assumption to derive the formula of Eq.~(\ref{eq:mix}) is to treat the neutrino component to stay completely 
at linear level. \cite{Saito:2009fk} shows that this assumption can be justified for expected small mass of neutrinos 
(see also \cite{Shoji:2010lr}). 
Given Eq.~(\ref{eq:mix}), \citet{Saito:2008lr,Saito:2009fk} proposed a SPT-based model to convert the matter power spectrum to the galaxy power spectrum at a given scale $k$ and a given redshift $z$, 
\begin{equation}\label{eq:nonlinearPkhalo}
  P_g(k; z) = b_{1}^{2}\left[P^{\rm 1loop}_{\rm cb\nu}(k; z)+b_{2}P_{\rm b2}(k;z)+b_{2}^{2}P_{\rm b22}(k;z)\right] + N,
\end{equation} 
where $b_1,b_2$ and $N$ denote the linear bias, nonlinear bias and the residual shot noise respectively, which can be derived using a SPT prescription \citep{McDonald:2006kx}. Quantities $P_{\rm b2},P_{\rm b22}$ can be calculated using SPT and the expression is explicitly given in Eqns.~(32) and~(36) in \cite{Saito:2009fk}. Note that on linear scales, Eq.~(\ref{eq:nonlinearPkhalo}) reduces to 
\begin{equation}\label{eq:blin} 
  P_g(k; z) = b_{1}^{2}P^{\rm L}_{\rm cb\nu}(k; z) + N.
\end{equation} 
Hence $b_1$ acts as a linear bias and $N$ contributes a shot noise contamination stemming from the stochastic bias and nonlinear clustering \citep{Heavens98,Seljak00,Smith07}. The terms multiplying $b_2$ give rise to a scale-dependent bias due to the nonlinear clustering. In general, $b_1,b_2,N$ vary with galaxy type, so we treat them as free parameters to be marginalised over in our analysis. 

\subsubsection{Modelling the redshift space distortions}

In this section we discuss the issue of model uncertainty in the CMASS power spectrum, 
focusing on the effect of the redshift-space distortions (RSD). 
In order to 
quantitatively model the impact of the RSD on the spherically averaged CMASS power spectrum, we compare the following RSD models:\\

\noindent{\it RSD Model 1: Linear Kaiser.} At very large scales where the linear perturbation theory 
holds, the mapping from real to redshift space can be expressed at linear order, and 
the resultant redshift-space power spectrum is enhanced by the so-called Kaiser factor \citep{Kaiser:1987vn}: 
\be
 P^{\rm S}_{g}(k, \mu) = (1+f\mu^{2})^{2}P_{g}(k), 
 \label{eq: Linear Kaiser}
\ee
where $\mu$ is the cosine between the line-of-sight direction and the wave vector, and $f$ is 
the logarithmic growth defined as $f\equiv d\ln D(a)/d\ln a$. The superscript `S' denotes the 
quantity in redshift space. Spherically averaging Eq.~(\ref{eq: Linear Kaiser}), we have the monopole 
component,  
\be
 P^{\rm S}_{0}(k) = \left(1+\frac{2f}{3}+\frac{f^{2}}{5}\right)P_{g}(k). 
\ee

\noindent{\it RSD Model 2: Nonlinear Kaiser.} 
The Linear Kaiser model is not true especially 
at mildly nonlinear scales (see \eg, \citealt{Scoccimarro:2004lr}). 
To linear order, the matter density perturbation in redshift space can be written as,
\be
 \delta^{\rm S}(k) = \delta(k) + f\mu^{2}\theta(k),
\ee
where $\theta$ is the divergence of the peculiar velocity field. 
In order to take into account nonlinear 
gravitational evolution of the velocity-divergence field separately, the Kaiser model is 
generalised to the {\it Nonlinear Kaiser} model as follows: 
\be
 P^{\rm S}_{g}(k, \mu) = P_{g,\delta\delta}(k) + 2f\mu^{2} P_{g,\delta\theta}(k) 
 + f^{2}\mu^{4}P_{\theta\theta}(k), 
 \label{eq: Nonlinear Kaiser}
\ee
where the galaxy density-density power spectrum, $P_{g,\delta\delta}(k)$ is 
modelled using Eq.~(\ref{eq:nonlinearPkhalo}), and the the galaxy density-velocity power 
spectrum is modelled as \citep{Swanson2010},
\be
 P_{g,\delta\theta}(k) = b_{1}\left[P_{\delta\theta}(k) + b_{2}P_{\rm b2,\theta}(k)\right]. 
\ee
Note that here we assume no velocity bias, and that the matter density-velocity, 
$P_{\delta\theta}$, or the velocity-velocity $P_{\theta\theta}$ can be computed 
using perturbation theory similarly to the density-density one. 
We compare SPT with the closure approximation (CLA) 
\citep{Taruya:2008lr, Nishimichi:2009uq} as an example. 
The CLA is one of the improved perturbation theories including the renormalized perturbation 
theory \citep{Crocce:2006uq, Crocce:2006qy}, and the CLA power spectrum at 2-loop order 
is in an excellent agreement with the $N$-body simulation results 
\citep{Taruya:2009sd, Carlson:2009ys}. A disadvantage of the CLA is that it involves 
time-consuming integrations in the 2-loop order, and therefore it is computationally difficult to apply the CLA 
to MCMC analysis (see \citealt{Taruya:2012lr} for recent effort to speed up the computation).\\

\noindent{\it RSD Model 3: Nonlinear Kaiser with FoG.}
At smaller scales than the typical size of virialized clusters, the internal velocity dispersion of 
galaxies makes the galaxy clustering pattern elongated along the line of sight, known as the 
Finger-of-God (FoG) effect \citep{Jackson:1972lr}. 
The FoG suppression 
is necessary for
massive halos in which most of the CMASS galaxies exist as central galaxies (\eg, \citealt{Hatton:1998lr}), with about $10\%$ of the CMASS galaxies being satellite galaxies \citep{White11BEDR}. The satellites are expected to have larger small-scale 
velocity dispersion than central galaxies, and cause the additional FoG suppression in the 
CMASS power spectrum. In order to account for the FoG effect, \cite{Scoccimarro:2004lr} 
proposed a phenomenological model in which the FoG suppression is described by an overall  
exponential factor: 
\ba
&&P^{\rm S}_{g}(k, \mu) = \exp\left(-f^{2}\sigma_{\rm V}^{2}k^{2}\mu^{2}\right)\nonumber\\
&&\;\;\;\;\;\; \times \left[P_{g,\delta\delta}(k) + 2f\mu^{2} P_{g,\delta\theta}(k) 
 + f^{2}\mu^{4}P_{\theta\theta}(k)\right], 
\label{eq: Nonlinear Kaiser + FoG}
\ea
where $\sigma_{\rm V}$ is the velocity dispersion which we treat as a free parameter. \\

\noindent{\it RSD Model 4: Nonlinear Kaiser with correction terms and FoG.} 
Recent studies show that higher-order correlations between the density and the velocity 
divergence in the nonlinear mapping from real to redshift space become important to 
explain the redshift-space power spectrum especially for massive halos \citep{Taruya:2010lr, 
Tang:2011lr, Nishimichi:2011lr, Okumura:2011fj, Reid:2011kx, Okumura:2012lr}. 
\cite{Taruya:2010lr} proposed a new model including such correction terms and can 
be generalized to biased objects:
\ba
&&P^{\rm S}_{g}(k, \mu) = \exp\left(-f^{2}\sigma_{\rm V}^{2}k^{2}\mu^{2}\right)\nonumber\\
&&\;\;\; \times \left[P_{g,\delta\delta}(k) + 2f\mu^{2} P_{g,\delta\theta}(k) 
 + f^{2}\mu^{4}P_{\theta\theta}(k)\right.\nonumber\\
&&\;\;\;\;\;\;\;\; \left.+ b_{1}^{3}A(k,\mu;\beta)+ b_{1}^{4}B(k,\mu;\beta)\right], 
\label{eq: Nonlinear Kaiser + AB + FoG}
\ea
with $\beta\equiv f/b_{1}$. Note that terms associated with $A$ and $B$ include the linear 
bias dependence of $b_{1}^{2}$ at maximum, and $b_{1}^{3}$ or $b_{1}^{4}$ originates 
from the fact that we replace $f$ with $\beta=f/b_{1}$. Also note that we did not include 
nonlinear bias terms proportional to $b_{2}$ in the correction terms for simplicity as we expect that such terms do not drastically affect the discussion here.  

\begin{figure}
\begin{center}
\includegraphics[width=0.45\textwidth]{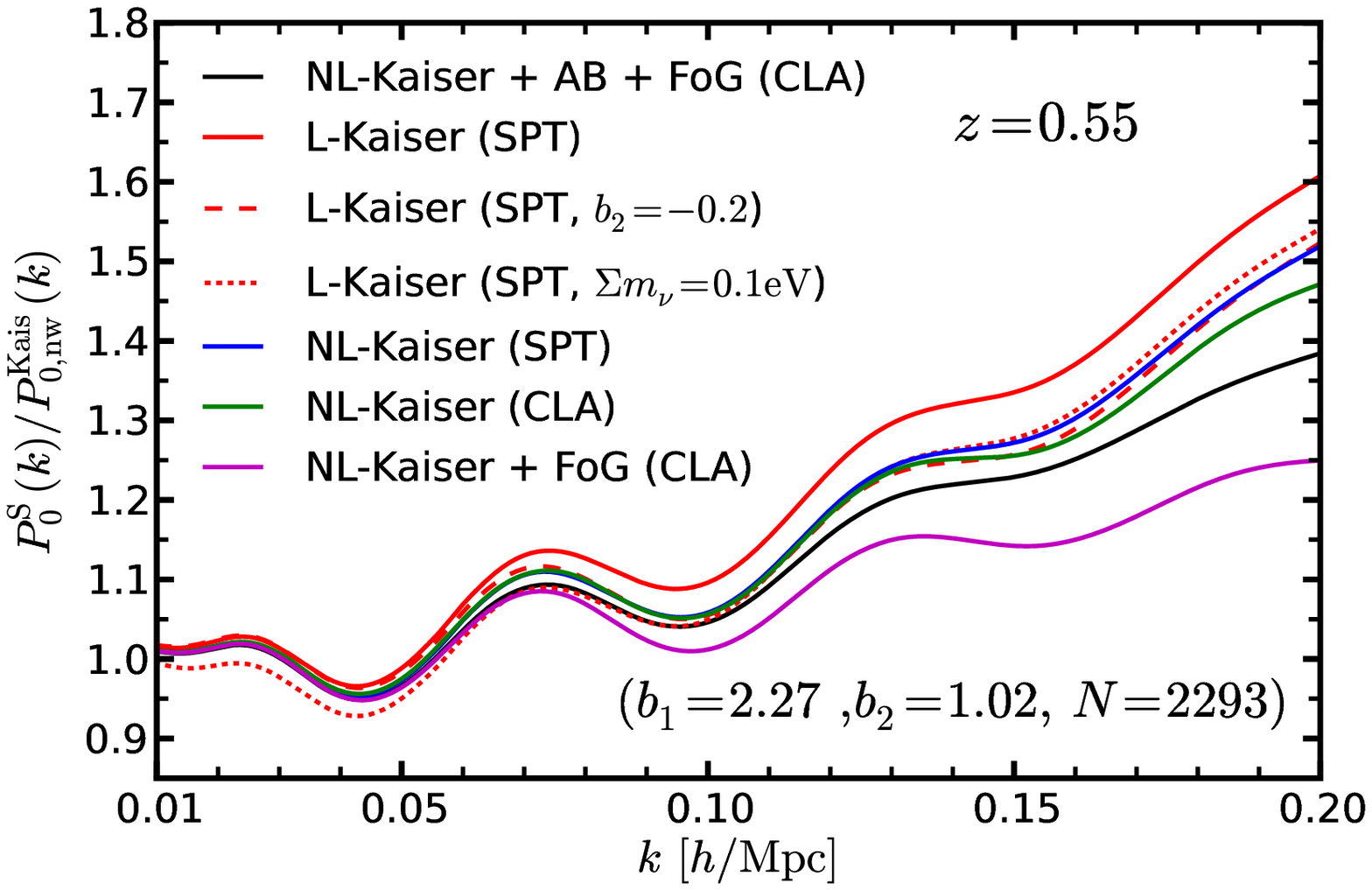}
\includegraphics[width=0.45\textwidth]{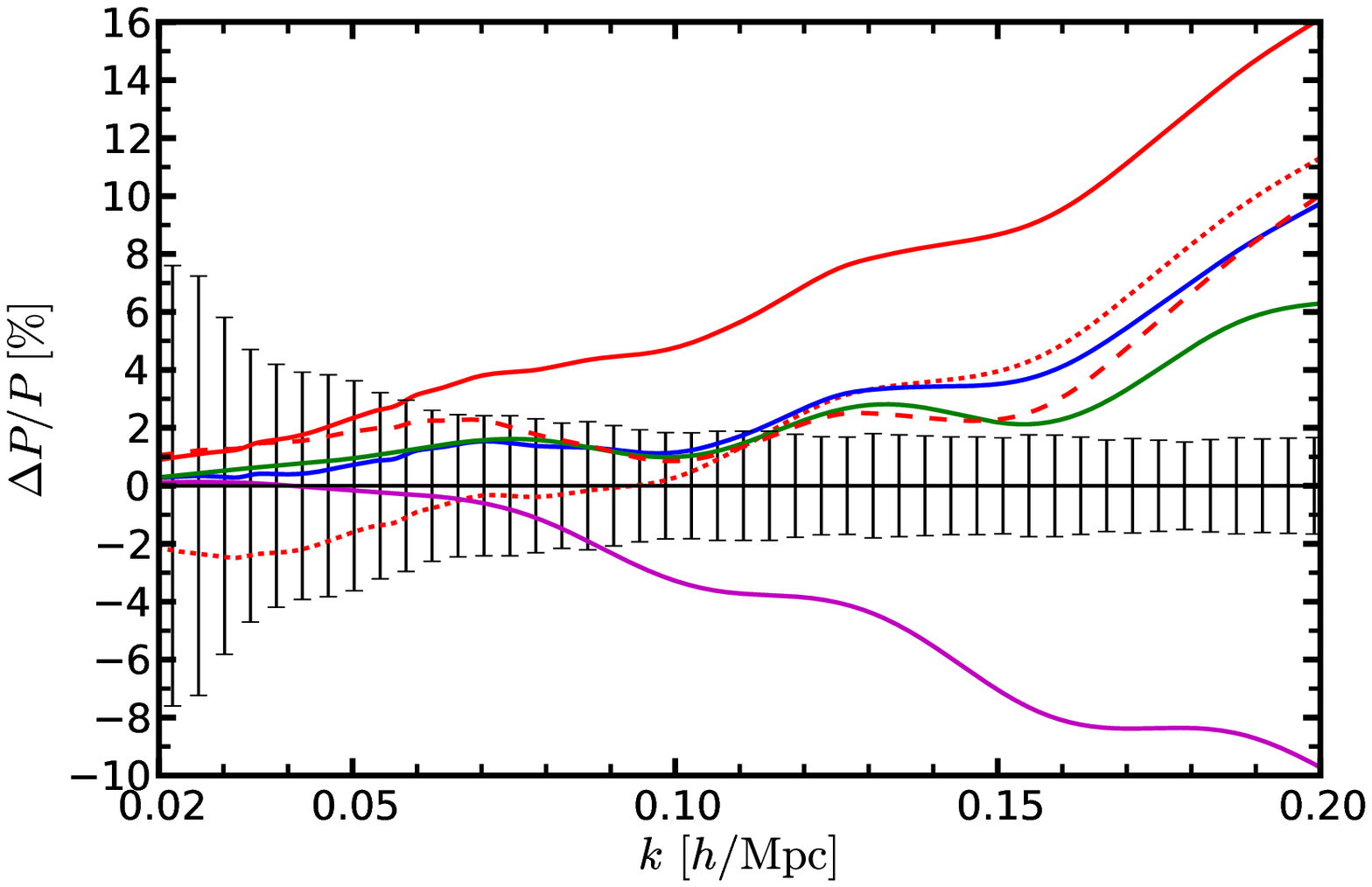}
\end{center}
\caption{An example of comparison among the RSD models. {\it Upper}: the monopole power spectra 
for the RSD models shown in the text; RSD 1: the Linear Kaiser ({\it red}), RSD 2: the Nonlinear Kaiser ({\it blue} 
for SPT and {\it green} for CLA), RSD 3: the Nonlinear Kaiser with the FoG prefactor ({\it magenta}), and 
RSD 4: the Nonlinear Kaiser plus correction terms with the FoG ({\it black}). 
Each spectrum is divided by the Linear Kaiser model with linear no-wiggle spectrum for clarification 
purpose. We consider the cosmology for the CMASS mocks and the best-fit parameters of 
$(b_{1},b_{2},N)$ in the case of $\Lambda m_{\nu}$CDM model. We use the linear velocity dispersion, 
$\sigma_{\rm V}=4.57\,{\rm Mpc}/h$ when computing the FoG prefactor. For comparison, the Linear Kaiser 
models with $b_{2}=-0.2$ ({\it red dashed}) and with $\sum m_{\nu}=0.1\,{\rm eV}$ ({\it red dotted}) are 
also shown. {\it Lower}: fractional difference of each model from the RSD model 4. The line colours 
and styles denote exactly same with those in the upper panel. We show the error bars taken from 
diagonal components in the CMASS covariance matrix as a reference.}
\label{fig: rsd}
\end{figure}


Now let us compare the predicted power spectra using the RSD models explained above, 
which are shown in Fig.~\ref{fig: rsd}. The upper panel shows absolute 
amplitudes of the monopole power spectrum (divided by the Linear Kaiser model with 
linear no-wiggle power spectrum from \citealt{Eisenstein:1998fj}) for each model. We set 
the values of bias parameters to the best-fit values in the case of $\Lambda m_{\nu}$CDM 
model, $(b_{1}=2.27,\,b_{2}=1.02,\,N=2293)$, corresponding to the Linear Kaiser 
for the RSD modeling. Meanwhile, the lower panel plots fractional differences from the model 
4 corresponding to the solid black line in the upper figure. As a reference we put the error 
bars taken from the diagonal components in the CMASS covariance matrix used in this paper. 
First of all, the Linear Kaiser using the SPT prediction (Model 1, red solid) overestimates 
the monopole amplitude at all scales compared to other models. The difference
between the Linear (Model 1, red solid) and the Nonlinear Kaiser (Model 2, blue or green) 
is typically $\sim 3\%$ at $k=0.1\,h/{\rm Mpc}$ or $\sim 6\%$ at $k=0.2\,h/{\rm Mpc}$ in the 
case of SPT. The Nonlinear Kaiser term does not strongly depend on the way $P_{\delta\delta}, P_{\delta\theta}$ and $P_{\theta\theta}$ are computed, \ie, with SPT or CLA 
if we consider $k<0.1\,h/{\rm Mpc}$. A comparison between the Nonlinear Kaiser (Model 2, green) 
with the Nonlinear Kaiser with FoG (Model 3, magenta) addresses how the FoG suppress 
the power spectrum. Here we choose $\sigma_{\rm V}=4.57\,{\rm Mpc}/h$ which corresponds 
to the value of linear velocity dispersion. Note that, since this value is expected to be larger 
than that of the real CMASS catalogues, the FoG suppression seen in the figure could be 
aggressive\footnote{This argument comes from the fact that almost all of the CMASS galaxies 
are central galaxies in the centre of massive halos and their velocity dispersions are 
expected to be small (\eg, \citealt{Nishimichi:2011lr}). Also, the effect of satellites on the 
monopole clustering is shown to be negligibly small, which is not true for the quadrupole 
(\citealt{ReidDR9,Okumura:2012lr})} .

In the specific case considered here, the FoG effect suppresses the amplitude by 
$\sim 5\%$ at $k=0.1\,h/{\rm Mpc}$ or $\sim 15\%$ at $k=0.2\,h/{\rm Mpc}$. 
Finally, since the correction terms $A$ and $B$ in Eq (\ref{eq: Nonlinear Kaiser + AB + FoG}) essentially enhance the amplitude at scales of 
interest, the RSD Model 4 prediction (black) becomes larger than that of the RSD Model 3 (magenta) 
by $\sim 4\%$ at $k=0.1\,h/{\rm Mpc}$ or $\sim 10\%$ at $k=0.2\,h/{\rm Mpc}$.\\

\subsection{Fitting formula}

\subsubsection{{\sc HALOFIT}-$\nu$}

Another approach to model the matter power spectrum in the weakly nonlinear regime is to use fitting formulae, such as {\sc HALOFIT} \citep{Smith2003}, calibrated using $N$-body simulations. {\sc HALOFIT} works well for models without massive neutrinos. However, \citet{Bird2012} found that {\sc HALOFIT} over-predicts the suppression of the power due to neutrino free-streaming on strongly nonlinear scales, and the discrepancy can reach the level of 10 percent at $k\sim1\hompc$. \citet{Bird2012} proposed a new fitting formula, {\sc HALOFIT}-$\nu$, which is an improved version of {\sc HALOFIT} when massive neutrinos are present. {\sc HALOFIT}-$\nu$ was calibrated using an extensive suite of $N$-body simulations with neutrinos and it was shown to work significantly better than the original {\sc HALOFIT} on nonlinear scales with the presence of neutrinos.      

To convert the matter power spectrum $P_{\rm HF\nu}$ calculated using {\sc HALOFIT}-$\nu$ into the observable galaxy power spectrum, we follow \cite{Swanson2010} combining this with free parameters for galaxy bias 
\begin{equation}\label{eq:bHF} 
  P_g(k; z) = b_{\rm HF}^{2}P_{\rm HF\nu}(k; z) + P_{\rm HF},
\end{equation} 
where $b_{\rm HF}$ and $P_{\rm HF}$ are constant model parameters representing the bias and the shot noise contamination respectively.  

\subsubsection{\citet{Cole2005}}

An alternative to model the galaxy power spectrum was proposed by \citet{Cole2005}, namely,
\begin{equation}\label{eq:bQ} 
  P_g(k; z) = b_{Q}^{2} \frac{1+Qk^2}{1+1.4k} P^{\rm L}_{\rm cb\nu}(k;z)
\end{equation} 
where $k$ is the wave number in unit of h/Mpc and the parameters $b_Q$ and $Q$ change the overall amplitude and the scale-dependence of the linear power spectrum on small scales. It was shown that this fitting formula can match the redshift-space galaxy power spectrum derived from the simulations well if the parameters $b,Q$ are properly chosen. In this analysis, we follow \citet{Swanson2010} to treat $b$ and $Q$ as free parameters to be marginalised over.    

\section{Other datasets and MCMC analysis}

To tighten the measurement of the summed neutrino mass, we combine the CMASS data with that from other cosmological surveys, including CMB, SN and Hubble parameter determinations.

\subsection{CMB and SN}

We use likelihoods based on the WMAP 7-year CMB data set including the temperature-temperature and temperature-polarisation angular power spectra \citep{Larson:2010gs}. 

There are two recently compiled, and publicly available sets of SN that are commonly used to set cosmological contraints -- the SNLS 3-year \citep{Conley:2011ku} and the Union2.1 sample \citep{Suzuki:2011hu}. We chose to use SNLS3 data in this work because it includes a more homogeneous sample of SN at higher redshift, which may make it less susceptible to systematics errors. We have also performed our analysis for some cases (\eg, weighing neutrino mass assuming a $\Lambda$CDM background cosmology) using Union2.1 data for a comparison (see Fig.~\ref{fig:omm_mnu_95cl}). For the SNLS data, the measured apparent magnitude $m_{\rm mod}$ is \citep{Conley:2011ku,Ruiz12}, 
\begin{equation}\label{eq:sn} 
  m_{\rm mod}=5{\rm log_{10}}\left(\frac{H_0}{c}d_L\right)-\alpha(s-1)+\beta \mathcal{C} + \mathcal{M},
\end{equation} 
where $d_L$ is the luminosity distance, and the presence of nuisance parameters $\alpha$ and $\beta$ allows the inclusion of the possible degeneracy between $m_{\rm mod}$, `stretch' $s$ and color $\mathcal{C}$. This is necessary because in principle the peak apparent magnitude of a SN Ia is degenerate with the broadness (stretch) of the light curve (broader is brighter), and with the colour as well (bluer is brighter) \citep{Ruiz12}. Thus in the likelihood calculation, where we essentially contrast the theoretical prediction to the measured $m_{\rm mod}$ for each SNe and calculate the usual quadratic sum of the signal-to-noise ratio, we numerically marginalise over $\alpha$ and $\beta$, and analytically marginalise over the constant magnitude offset $\mathcal{M}$. 

\subsection{BAO}

\begin{table}

\begin{center}
\begin{tabular}{cccc}

\hline\hline

               		                 & $z$           &$r_s/D_V$                             & $A(z)$      \\ \hline
6dF        				&$0.106$     & $0.336\pm0.015$                 & $-$           \\ \hline
\multirow{2}{*}{SDSS-II}    	& $0.20$      & $0.1905\pm0.0061$             & $-$           \\
                                          	& $0.35$      & $0.1097\pm0.0036$             & $-$           \\  \hline
\multirow{3}{*}{WiggleZ}   	& $0.44$      &   $-$                                      & $0.474\pm0.034$        \\
                                       	& $0.60$      &   $-$                                      & $0.442\pm0.020$        \\
                                          & $0.73$      &   $-$                                      & $0.424\pm0.021$        \\ \hline
BOSS      	                		&$0.57$       &    $0.07315\pm0.00118$       & $-$           \\                                          
\hline\hline

\end{tabular}
\end{center}
\caption{The BAO measurements used in this work. The quantity $r_s/D_V$ is the ratio of the sound horizon $r_s$ to $D_V$ defined in Eq (\ref{eq:DV}), and $A(z)$ is the parameter defined in \citet{Eisenstein05}.}
\label{tab:bao}
\end{table}%

\begin{figure*}
\begin{center}
\includegraphics[scale=0.2]{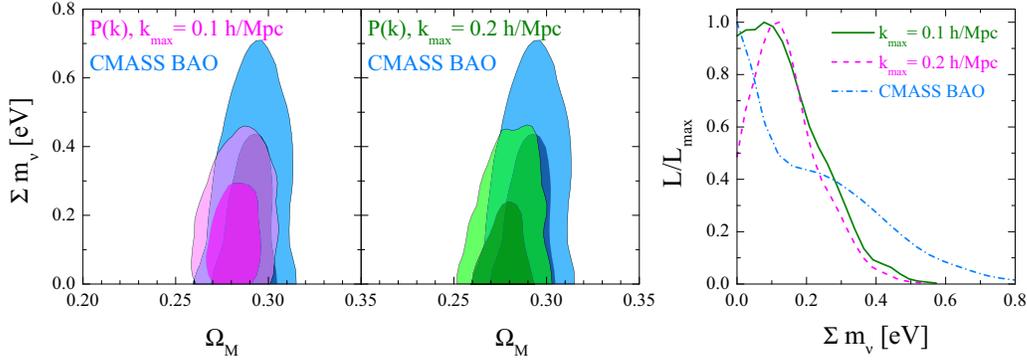}
\end{center}
\caption{The purple and green contours on the top layer in left panels: the 68 and 95 percent CL contour plots for neutrino mass and $\Omega_{\rm M}$ obtained from the joint dataset including CMB+SN+CMASS power spectra cut off at various $k$ illustrated in the figure; The blue contours on the bottom layer in left panels: the 68 and 95 percent CL contour plots for neutrino mass and $\Omega_{\rm M}$ obtained from the joint dataset including CMB+SN+CMASS BAO; Right panel: the corresponding $1D$ posterior distribution of neutrino mass. A $\Lambda$CDM model is assumed for the background cosmology.
}\label{fig:mnu_kmax}
\end{figure*}

We also combine the baryonic acoustic oscillations (BAO) measurements from 6dF \citep{6df}, SDSS-II \citep{Percival10} and the WiggleZ survey \citep{wigglez}. The data used is shown in Table \ref{tab:bao}. Since the measurements of CMASS $P(k)$ and BOSS BAO \citep{alph} use the same galaxy sample, we never use the $P(k)$ and BAO measurements of BOSS simultaneously. But we shall compare the neutrino mass constraint using BOSS $P(k)$ and BAO respectively (see Fig.~\ref{fig:mnu_kmax} and later discussions). The difference in the constraints will quantify the degree to which it is the growth supression effect of massive neutrinos that provides the information we recover.

\subsection{Parameterising the Universe and the MCMC engine}

The set of cosmological parameters that we will simultaneously measure is given in Table~\ref{tab:param}, which also provides the basic definitions, the prior assumed, and the values for the vanilla $\Lambda$CDM model for these parameters. We use $\Theta_s$ rather than $H_0$ to parameterise the background expansion of the universe since $\Theta_s$ is less degenerate with other parameters \citep{Lewis:2002ah}. A few require lengthier definitions, including $w_0$ and $w_a$, which parameterise the equation-of-state (EoS) parameter $w$  of dark energy as a function of the scale factor $a$
\begin{equation}\label{eq:CPL} 
  w(a)=w_0+w_a(1-a).
\end{equation} 
This is the so-called CPL parametrisation proposed by \citet{CP} and \citet{Linder}. Parameters $A_s,n_s$ and $\alpha_s$ parameterise the primordial power spectrum $P_{\rm in}$ \citep{running}, 
\begin{equation} 
  {\rm ln}P_{\rm in}(k)\equiv {\rm ln}A_s + [n_s(k_{\rm pv})-1]{\rm ln}
    \left(\frac{k}{k_{\rm pv}} \right)+\frac{\alpha_s}{2}\left[{\rm ln}\left(\frac{k}{k_{\rm pv}}\right) \right]^2,
\end{equation} 
where $k_{\rm pv}$ denotes the pivot scale for the parametrisation and we choose $k_{\rm pv}=0.05\hompc$ in this work.

We use a modified version of {\sc CAMB} \citep{CAMB} to calculate the observables and compute the likelihood by comparing to observations. In our general parameter space (see Table~\ref{tab:param}), $w(a)$ evolves with the scale factor $a$, and it is allowed to cross $w=-1$. This means that we allow the quintessence \citep{quintessence1,quintessence2}, phantom \citep{phantom}, k-essence \citep{kessence} and quintom \citep{quintom} dark energy models in the global fitting. Note that when $w(a)$ crosses the divide, \ie, 
\begin{eqnarray}\label{eq:quintom} 
  ~~~w_0<-1~~{\rm and}~~w_0+w_a>-1,&& \\ \nonumber 
  ~~~{\rm or}~~w_0>-1~~{\rm and}~~w_0+w_a<-1,&&
\end{eqnarray}
dark energy fluid has multi-components \citep{quintom,Hu05}. In this case, we follow the prescription proposed by \citet{DEP} to calculate the evolution of dark energy perturbations \footnote{Note that alternative approaches have been proposed and can yield consistent result (\eg, see~\citealt{Fang08}).}. We use a modified version of {\sc CosmoMC} \citep{Lewis:2002ah}, which is a Monte Carlo Markov Chain (MCMC) engine, to efficiently explore the multi-dimensional parameter space using the Metropolis-Hastings algorithm \citep{Metropolis53,Hastings70}. 

To summarise our default assumptions, unless specifically mentioned, 
\begin{itemize}
\item  we use Eq.~(\ref{eq:nonlinearPkhalo}) to model the galaxy power spectrum;
\item  we use WMAP seven-year power spectra and the SNLS three-year measurement for the CMB and SN data respectively;
\item  we use the CMASS galaxy power spectrum measurement truncated at $k=0.1\hompc$ for the `CMASS' data;
\item  when using the `CMASS' data, we always combine the BAO measurement shown in Table \ref{tab:bao} {\it except for} the BOSS measurement;
\item  all the neutrino mass constraints are the 95 percent CL upper limit;
\item  all $P(k)$ shown are evaluated at the mean redshift of BOSS, which is $z_{\rm eff}=0.57$.
\end{itemize}

\begin{figure}
\begin{center}
\includegraphics[scale=0.15]{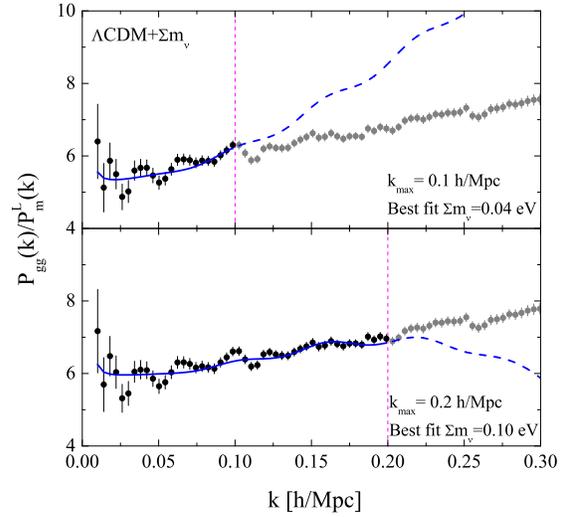}
\end{center}
\caption{The CMASS data used in the analysis and the best fit power spectrum assuming a $\Lambda m_{\nu}$CDM cosmology. The data and spectra are both rescaled using the linear matter spectrum for the best fit model. The upper and lower panels show the cases of $k_{\rm max}=0.1$ and $0.2\hompc$ respectively. }\label{fig:pkfit}
\end{figure}
\section{Results}

\subsection{Neutrino mass measurement in a $\Lambda$CDM universe}

We now measure the summed neutrino mass $\Sigma m_{\nu}$ assuming that the background cosmology follows that of a $\Lambda$CDM model. \ie, only the {\it vanilla} cosmological parameters $\omega_b$, $\omega_c$, $\Theta_s$, $\tau$, $n_s$, $A_s$, $\Sigma m_{\nu}$, and the nuisance parameters are allowed to vary. 

\subsubsection{The choice of $k_{\rm max}$}

Fitting to a wider range of scales has the potential to provide better constraints on $\Sigma m_{\nu}$, as we have a longer lever-arm with which to assess the contribution from the neutrinos. However fitting to smaller scales relies more heavily on the model for galaxy formation. In Fig.~\ref{fig:mnu_kmax}, we compare results fitting to scales with $k_{\rm max}=0.1,0.2\hompc$ when all data are combined (purple and green contours). As shown, the change is marginal, with the upper limit for $\Sigma m_{\nu}$ only lowered to $0.338$\,eV from $0.340$\,eV when $k_{\rm max}$ is increased from $0.1$ to $0.2 \hompc$. This is understandable given our default galaxy bias model: when $k_{\rm max}$ is larger, the non-linear $P(k)$ data simply constrains the nuisance parameters $b_2$ in Eq.~(\ref{eq:nonlinearPkhalo}) rather than $\Sigma m_{\nu}$. We should expect that, for larger $k_{\rm max}$, the model Eq.~(\ref{eq:nonlinearPkhalo}) becomes less reliable since it is based on perturbation theory. Given that our lack of knowledge of the non-linear bias of the CMASS galaxies means there is little information over the range $0.1<k<0.2\hompc$, assuming our model is appropriate on these scales is not worth the added risk, and we choose $k_{\rm max}=0.1 \hompc$ as our default. All of the results presented are based on this conservative limit.

Fig.~\ref{fig:pkfit} shows the goodness-of-fit by over-plotting the best-fit model on top of the observational data for the cases of $k_{\rm max}=0.1,0.2 \hompc$. The quantity shown is the ratio between the galaxy power spectrum and the linear matter power spectrum, $P_g(k)/P^{\rm L}_{\rm cb\nu}$ (see Eqns.~(\ref{eq:mix}) and~(\ref{eq:nonlinearPkhalo})). From Eq.~(\ref{eq:blin}), we can see that on large scales, this ratio is roughly the linear galaxy bias squared $b_1^2$. Our MCMC analysis suggests $b_1\simeq2$, which is consistent with the result in \citet{ReidDR9}.

\begin{figure*}
\begin{center}
\includegraphics[scale=0.5]{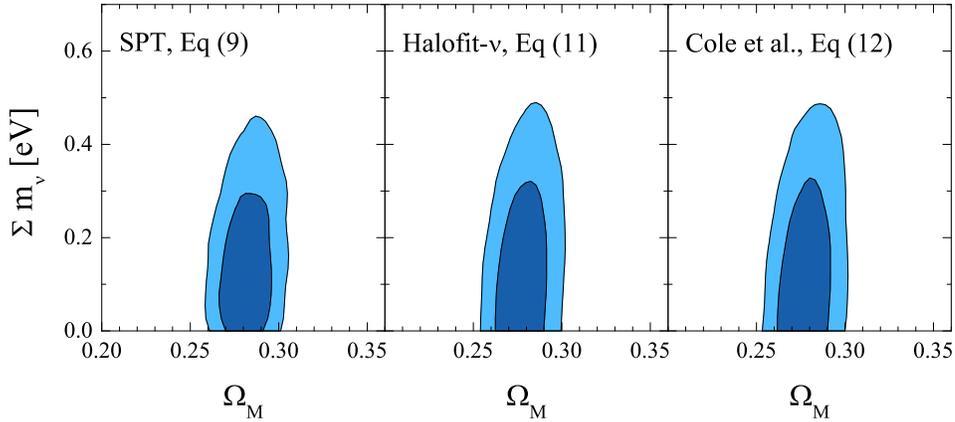}
\end{center}
\caption{The 68 and 95 percent CL contour plot for neutrino mass and $\Omega_{\rm M}$ using three different galaxy modelling.}\label{fig:mnu_PTHalofit}
\end{figure*}

\subsubsection{The choice of the galaxy modelling}

Next we shall test the effect of the choice of the galaxy modelling. To test the limiting scales to be fitted in the last section, we used Eq.~(\ref{eq:nonlinearPkhalo}) to model the galaxy power spectrum. In Fig.~\ref{fig:mnu_PTHalofit}, we show the contours for $\Sigma m_{\nu}$ and $\Omega_{\rm M}$ for the {\sc HALOFIT}-$\nu$ (Eq.~\ref{eq:bHF}) and the \citet{Cole2005} (Eq.~\ref{eq:bQ}) models. The difference between constraints calculated assuming these three models is marginal, which is reasonable as they only differ in form for $k_{\rm max}>0.1\hompc$.

\subsubsection{The choice of SN data}

\begin{figure}
\begin{center}
\includegraphics[scale=0.17]{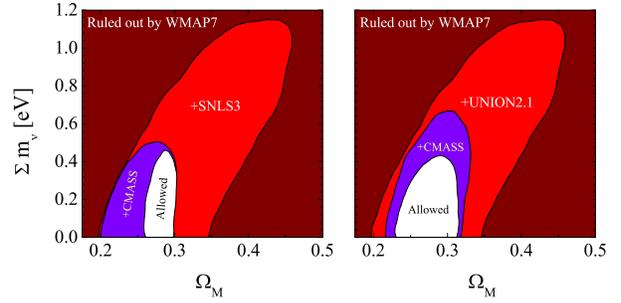}
\end{center}
\caption{The 95 percent CL allowed parameter space for $(\Omega_{\rm M},\Sigma m_{\nu})$ obtained from the joint dataset including CMB (red), CMB+SN(purple), and CMB+SN+CMASS $P(k)$ (white) in a  $\Lambda m_{\nu}$CDM model. The SN data of SNLS 3-year and the Union2.1 sample were used for the left and right panels respectively. The CMASS $P(k)$ was truncated at $k_{\rm max}=0.1\hompc$.}\label{fig:omm_mnu_95cl}
\end{figure}

The left panel of Fig.~\ref{fig:omm_mnu_95cl} shows 95 percent CL contour plots for $\Sigma m_{\nu}$ and $\Omega_{\rm M}$, comparing various data combinations. Comparison with the left panel shows the effect of including the Union2.1 rather than SNLS3 SN data.  SNLS3 data provides tighter constraint on $\Omega_{\rm M}$, although the measurements of the neutrino mass are similar, namely,  
\begin{eqnarray} \label{eq:kmax}
&&\Sigma m_{\nu}<0.340~{\rm eV~(WMAP7+SNLS3+CMASS)}, \nonumber\\
&&\Sigma m_{\nu}<0.334~{\rm eV~(WMAP7+Union2.1+CMASS)}.
\end{eqnarray}
In both cases, we see that CMASS data help to reduce the allowed parameter space dramatically. 

\subsubsection{The effect of the redshift space distortion modelling}
\label{sec:RSD}

\begin{figure}
\begin{center}
\includegraphics[scale=0.3]{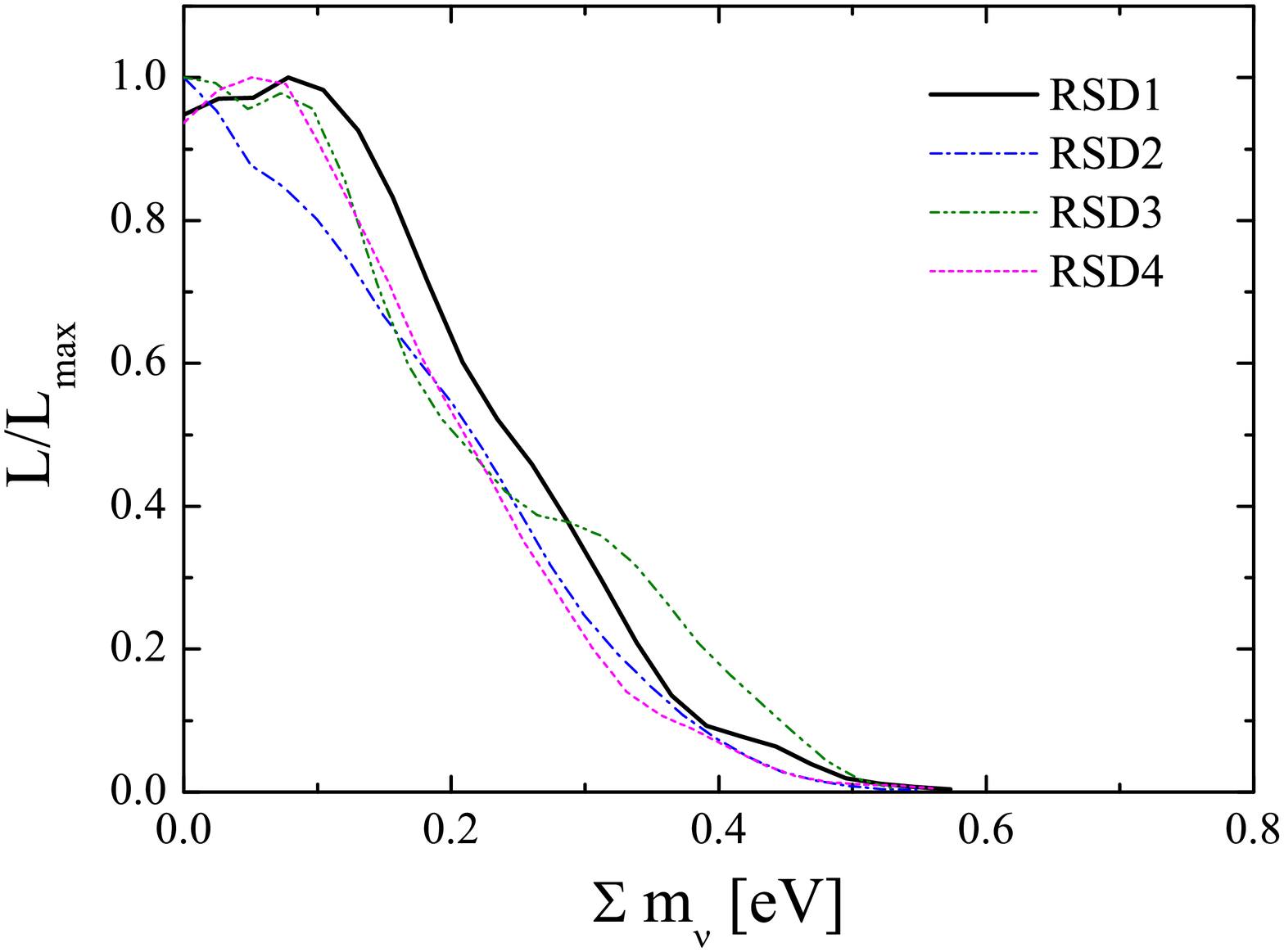}
\end{center}
\caption{The 1-D posterior distribution of the neutrino mass using different RSD modelling as illustrated in the legend.}\label{fig:RSD_1D}
\end{figure}

\begin{figure*}
\begin{center}
\includegraphics[scale=0.35]{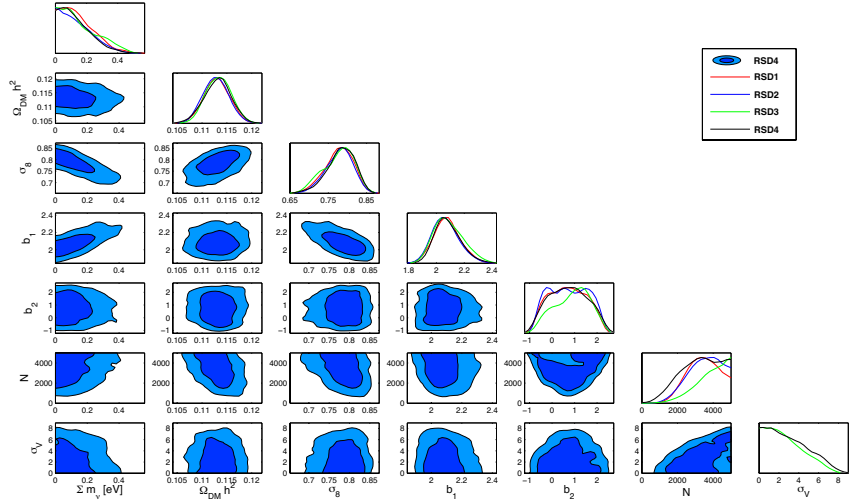}
\end{center}
\caption{The 1-D and 2-D constraint on neutrino mass and related parameters.}\label{fig:RSD_tri}
\end{figure*}

We now test how the neutrino mass measurements are affected by the choice of the RSD model. In Figs \ref{fig:RSD_1D}, \ref{fig:RSD_tri} and Table \ref{tab:RSD}, we show the constraint on neutrino mass (and related parameters) derived from the same dataset (the CMASS $P(k)$ is cut at $k=0.1 h$/Mpc) using different RSD modelling. As we can see, the neutrino mass constraints are generally similar. This is mainly because the effect of scale-dependent RSD up to $k=0.1~h$/Mpc is rather mild and is unlikely to be degenerate with the neutrino mass. Note, however, the neutrino mass is indeed degenerate with the linear bias parameter $b_1$ with the correlation coefficient being $0.716$. This is easy to understand: increasing $b_1$ shifts the whole $P(k)$ upwards, thus a larger neutrino mass is needed to suppress the power to compensate. This is similar to the well-known degeneracy between the neutrino mass and the equation-of-state $w$ of dark energy \citep{Hannestad05}. 

Given these results, in this work we adopt the RSD1 model, which is equivalent to the SPT model (Eq.~\ref{eq:nonlinearPkhalo}), as a default just for simplicity.


\begin{table}

\begin{center}
\begin{tabular}{c|c|c|c}

\hline\hline
&  \multicolumn{3}{c}{$\Sigma m_{\nu}$ [eV]} \\
RSD modelling&   95 percent CL & Mean & $\sigma$\\ \hline
RSD1 &  $<0.340$& $0.149$ & $0.104$\\
RSD2 &  $<0.336$& $0.140$ & $0.103$\\
RSD3 &  $<0.384$& $0.158$ & $0.118$\\
RSD4 &  $<0.324$& $0.135$ & $0.099$\\

\hline\hline

\end{tabular}
\end{center}
\caption{The neutrino mass constraint for various RSD modelling.}
\label{tab:RSD}
\end{table}%

\subsubsection{$P(k)$ or BAO?}

A strong feature of the galaxy power spectrum is the BAO signal, whose location depends on $r_s/D_V$, where $r_s$ is the sound horizon at the baryon drag epoch, and $D_V$ is the average distance to the galaxies pairs defined in Eq~(\ref{eq:DV}). 
The full CMASS $P(k)$ includes information on additional physical processes, including those related to neutrinos. In some cases, \eg, for dark energy parameters, BAO and the full $P(k)$ can provide similar constraints (see, \eg, \citealt{Sanchez:2012sg}). By comparing our constraints using the full $P(k)$ to the BAO only result, we can determine the amount of information about physical processes related to neutrinos that is encoded in the CMASS $P(k)$.   

For the constraints on the summed neutrino mass, Fig.~\ref{fig:mnu_kmax} and Table~\ref{tab:kmax}, show that the constraint from including the CMASS BAO measurement, rather than the full power spectrum is much weaker, namely, 
\begin{eqnarray}
  &&\Sigma m_{\nu} < 0.579~{\rm eV~(WMAP7+SNLS3+CMASS~BAO)}, \nonumber  \\ 
  &&\Sigma m_{\nu} < 0.340~{\rm eV~(WMAP7+SNLS3+CMASS}~P(k)). \nonumber 
\end{eqnarray} 
Clearly, the broadband shape of the CMASS $P(k)$ contains significant information on physical processes related to neutrinos, particularly on the small-scale damping of $P(k)$ caused by the free-streaming of neutrinos. 

\subsubsection{Summary}

\begin{table*}

\begin{center}
\begin{tabular}{c|c|c|c|c}

\hline\hline
& & \multicolumn{3}{c}{$\Sigma m_{\nu}$ [eV]} \\
Galaxy modelling& Data &  95 percent CL & Mean & $\sigma$\\ \hline
SPT, Eq.~(\ref{eq:nonlinearPkhalo}) &CMASS $P(k), k_{\rm max}=0.1\hompc$ & $<0.340$& $0.149$  & $0.104$\\
SPT, Eq.~(\ref{eq:nonlinearPkhalo}) & CMASS $P(k), k_{\rm max}=0.2\hompc$ &  $<0.338$& $0.152$ & $0.097$\\
{\sc HALOFIT}, Eq.~(\ref{eq:bHF}) & CMASS $P(k), k_{\rm max}=0.1\hompc$ & $<0.388$& $0.167$  & $0.117$\\
{\sc HALOFIT}, Eq.~(\ref{eq:bHF}) & CMASS $P(k), k_{\rm max}=0.2\hompc$ & $<0.402$&  $0.173$ &  $0.121$\\
\citet{Cole2005}, Eq.~(\ref{eq:bQ}) & CMASS $P(k), k_{\rm max}=0.1\hompc$ & $<0.399$&  $0.168$ &  $0.121$\\
$-$ & CMASS BAO &  $<0.579$ & $0.261$   & $0.172$\\
\hline\hline

\end{tabular}
\end{center}
\caption{The neutrino mass constraint for various galaxy modelling and data choices.}
\label{tab:kmax}
\end{table*}%

The measurements of the neutrino mass in a $\Lambda$CDM background cosmology are summarised in Table \ref{tab:kmax}. We find that 
\begin{enumerate}
  \item As long as WMAP7 and CMASS $P(k)$ data are used, using SNLS3 or Union2.1 SN data yield similar neutrino mass constraints;
  \item The neutrino mass constraint cannot be improved when fitting to scales $k_{\rm max}\gtrsim 0.1\hompc$;
  \item If $k_{\rm max}=0.1 \hompc$, three galaxy models for the galaxy power spectrum described in  Eqns.~(\ref{eq:nonlinearPkhalo}), (\ref{eq:bHF}) and~(\ref{eq:bQ}), and the 4 models
for redshift space distortions described in section \ref{sec:PT} yield similar results;
  \item The constraint is significantly degraded if the BAO only are used instead of the full power spectrum.
\end{enumerate}

\subsection{Neutrino mass measurements in more general background cosmologies}

\begin{table}

\begin{center}
\begin{tabular}{c|c}

\hline\hline
                             
            Floating parameters                      &  Acronym   \\ \hline
                                                           
 $\mathcal{V}$ &   $\Lambda$CDM      \\                                
 $\mathcal{V}+\Sigma m_{\nu}$&  $\Lambda m_{\nu}$CDM \\
 $\mathcal{V}+\Sigma m_{\nu} + N_{\rm eff}$ &  $\Lambda N_{\rm eff} m_{\nu}$CDM   \\
 $\mathcal{V}+\Sigma m_{\nu} + w_0$ &  $w m_{\nu}$CDM   \\
 $\mathcal{V}+\Sigma m_{\nu} + w_0 + w_a$ &  $w_0 w_a m_{\nu}$CDM   \\
  $\mathcal{V}+\Sigma m_{\nu} + \alpha_s$ &  $\Lambda m_{\nu} \alpha_s$CDM   \\
 $\mathcal{V}+w_0$&  $w$CDM     \\
  $\mathcal{V}+w_0+w_a$&  $w_0 w_a$CDM     \\
 $\mathcal{V}+\Omega_K$&   $o$CDM   \\
 $\mathcal{V}+\alpha_s$&       $\Lambda\alpha_s$CDM  \\   
 $\mathcal{V}+r$      & $\Lambda r$CDM   \\
$\mathcal{V}+\Sigma m_{\nu} + N_{\rm eff}+w_0+w_a+\Omega_K+r+\alpha_s$ & All float\\                                    

\hline\hline
\end{tabular}
\end{center}
\caption{The floating cosmological parameters and the corresponding acronym for each models. The symbol $\mathcal{V}$ stands for the set of default vanilla parameters defined in Table~\ref{tab:param}.}
\label{tab:acronym}
\end{table}%

In this section, we reinvestigate the summed neutrino mass measurement in more general cosmologies, when several parameters in Table~\ref{tab:param} are allowed to vary. Figs.~\ref{fig:omm_mnu_models} and~\ref{fig:mnu_1D_models} present likelihood contours showing neutrino mass constraints in the models considered

\subsubsection{Neutrino mass in a $w m_{\nu}$CDM universe}

\begin{figure}
\begin{center}
\includegraphics[scale=0.3]{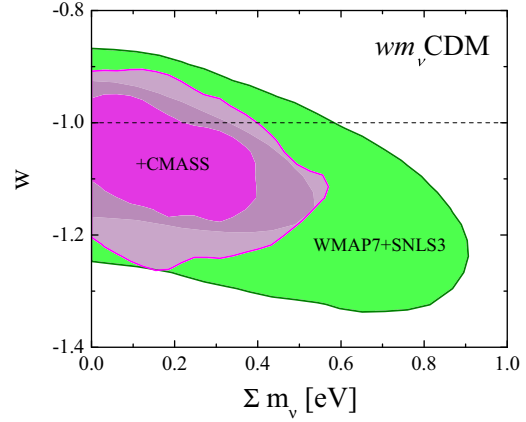}
\end{center}
\caption{The 68 and 95 percent CL contour plots for $w$ and neutrino mass obtained from the joint dataset including CMB+SN (green) and CMB+SN+CMASS power spectrum (magenta) in a $w m_{\nu}$CDM cosmology. }\label{fig:w_mnu}
\end{figure}

Neutrino mass and $w$ (assumed to be a constant here) of dark energy are generally degenerate because both can affect the shape of the matter and CMB power spectra \citep{Hannestad05}. We indeed see this degeneracy in Fig.~\ref{fig:w_mnu}, and this is why the neutrino mass constraint is diluted if $w$ is allowed to float, as shown in panel (A) in Fig.~\ref{fig:omm_mnu_models}. In Table~\ref{tab:mnu_models}, we can see that when $w$ is varied, the 95 percent CL upper limit for neutrino mass is relaxed from $0.340$ to $0.432$\,eV, a 27 percent degradation. 

\subsubsection{Neutrino mass in a $w_0w_a m_{\nu}$CDM universe}

\begin{figure}
\begin{center}
\includegraphics[scale=0.14]{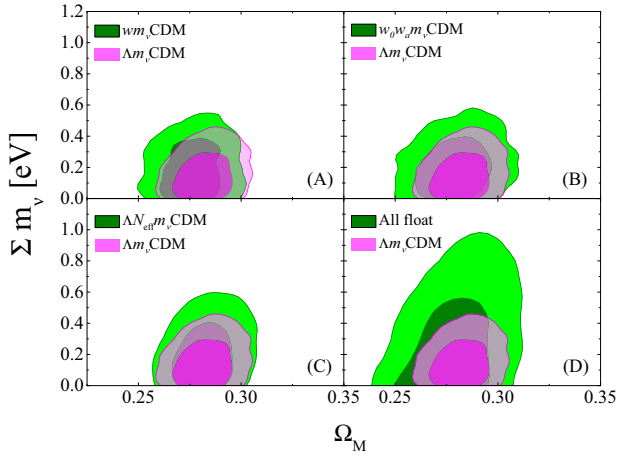}
\end{center}
\caption{Green contours in the back layer: the 68 and 95 percent CL constraints for the summed neutrino mass and $\Omega_{\rm M}$ assuming different cosmologies shown in the legend. Transparent magenta contours on the front layer: the constraint assuming a $\Lambda m_{\nu}$CDM cosmology for a comparison. }\label{fig:omm_mnu_models}
\end{figure}

\begin{figure}
\begin{center}
\includegraphics[scale=0.25]{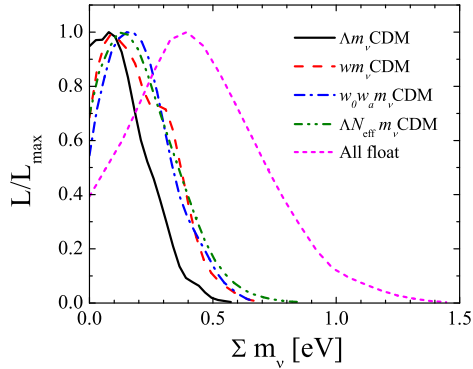}
\end{center}
\caption{The corresponding $1D$ posterior distribution of neutrino mass for the cases shown in Fig.~\ref{fig:omm_mnu_models}. }\label{fig:mnu_1D_models}
\end{figure}

The dynamics of dark energy can further dilute the neutrino mass constraint \citep{Xia07}. We measure the neutrino mass again by parameterising dark energy using the CPL parametrisation, Eq.~(\ref{eq:CPL}). As shown in panel (B) of Figs.~\ref{fig:omm_mnu_models} and~\ref{fig:mnu_1D_models} and Table~\ref{tab:mnu_models}, the neutrino mass constraint is further diluted to $0.465$\,eV, which is a 37 percent degradation.

\subsubsection{Neutrino mass in a $\Lambda N_{\rm eff} m_{\nu}$CDM universe}

The effective number of the relativistic species $N_{\rm eff}$ is directly related to $\Omega_{\rm M} h^2$, the physical matter density, and $z_{\rm eq}$, the redshift of the matter-radiation equality. Consequently, varying $N_{\rm eff}$ can change the ratio of the first to the third peak of the CMB temperature-temperature spectrum, and also the shape of matter power spectrum \citep{Komatsu11}. The standard value of $N_{\rm eff}$ is $3.046$ \citep{Mangano05}, but in general, $N_{\rm eff}$ can be considered a free parameter to be constrained by data. The bound of $N_{\rm eff}$ can be used to study the neutrino physics, and investigate the upper bound of energy density in primordial gravitational waves with frequencies $>10^{15}$ Hz \citep{Komatsu11}. As we see from panel (C) in Fig.~\ref{fig:omm_mnu_models} and in Table~\ref{tab:mnu_models}, relaxing $N_{\rm eff}$ dilutes the neutrino mass more than the dark energy parameters: the constraint is relaxed to $0.491$\,eV, which is a 44 percent degradation. 

\subsubsection{Neutrino mass in the most general cosmology}

Finally, let us consider the most conservative case in which all the cosmological parameters listed in Table~\ref{tab:param} are allowed to vary simultaneously: in addition to $w_0$, $w_a$, \&~$N_{\rm eff}$, the curvature $\Omega_K$, running $\alpha_s$ and tensor-to-scalar ratio $r$ are varied. All of these parameters can, in principle, be degenerate with the neutrino mass, \eg, $\alpha_s$ can mimic the effect of neutrinos on $P(k)$ by titling the power, thus changing the shape \citep{Feng06}. As expected, the neutrino mass constraint is weakened significantly to $0.879$\,eV, a $160$ percent degradation compared to that in a $\Lambda$CDM background cosmology.

\subsection{Measurement of other cosmological parameters}

\begin{table*}

\begin{center}
\begin{tabular}{c|c|c|c|c|c|c|c}

\hline\hline
                                    & $\Lambda$CDM     &    $\Lambda m_{\nu}$CDM & $\Lambda m_{\nu}N_{\rm eff}$CDM   & $w m_{\nu}$CDM &$w_0 w_a m_{\nu}$CDM & $\Lambda m_{\nu}\alpha_s$CDM & All float \\ \hline
                                    
$\Sigma m_{\nu}$ [eV]    &0            & $<0.340$            &      $<0.491$    &  $<0.432$  &  $<0.618$ &  $<0.395$  &$<0.821$  \\
$N_{\rm eff}$                   &3.046     &   3.046           &     $4.308\pm0.794$        &  3.046  & 3.046  &    3.046 &$4.032^{+0.870}_{-0.894}$    \\
$w_0$                              &        $-1$      &     $-1$         &       $-1$     &  $-1.081\pm0.075$  &  $-0.982^{+0.157}_{-0.156}$ &  $-1$   &$-0.964^{+0.173}_{-0.168}$    \\
$w_a$                              &          0    &       0       &         0     &   0 &  $-0.718^{+0.911}_{-0.966}$ &   0  &$-0.731^{+0.982}_{-1.053}$  \\
$100\Omega_K$               &      0           &      0        &        0     &  0  &   0 &   0  &$-0.411\pm1.02$   \\
$r$                   &        0      &       0       &        0       &   0 &  0 &  0  &$<0.440$   \\
$\alpha_s$       &     0         &       0     &       0       &   0  &  0 &  $-0.012\pm0.019$ & $-0.032\pm0.030$   \\       \hline                             
                                    
$100\Omega_b h^2$   &      $2.265\pm0.050$                  &    $2.278\pm0.054$         &       $2.261\pm0.051$       &  $2.268^{+0.050}_{-0.052}$  &  $2.241^{+0.053}_{-0.052}$ & $2.246\pm0.060$ &   $2.268^{+0.080}_{-0.081}$    \\
$100\Omega_c h^2$   &      $11.317^{+0.243}_{-0.245}$    &    $11.217^{+0.254}_{-0.256}$          &    $13.31^{+1.32}_{-1.30}$        & $11.455^{+0.352}_{-0.341}$   &  $11.765\pm0.472$ &   $11.303^{+0.264}_{-0.262}$    & $12.974^{+1.304}_{-1.329}$  \\
$\Theta_s$                  &     $1.0397^{+0.0024}_{-0.0025}$&    $1.0394^{+0.0026}_{-0.0027}$          &     $1.0350\pm{0.0038}$       &  $1.0390\pm{0.0026}$   & $1.0387\pm0.0026$  &  $1.0395\pm0.0026$   &   $1.0364\pm0.0039$  \\
$\tau$                          &      $0.0870^{+0.0063}_{-0.0070}$&     $0.0869^{+0.0065}_{-0.0069}$         &      $0.0881^{+0.0063}_{-0.0069}$       &  $0.0854^{+0.0061}_{-0.0063}$  & $0.0850^{+0.0061}_{-0.0068}$ & $0.0921^{+0.0071}_{-0.0080}$ &  $0.0927^{+0.0069}_{-0.0082}$       \\
$n_s$                          &       $0.967\pm0.012$                    &      $0.969\pm0.012$        &      $0.982\pm0.014$      & $0.966\pm0.012$   & $0.961\pm0.013$  &    $0.954\pm0.026$   & $0.965^{+0.043}_{-0.042}$ \\
ln$[10^{10} A_s]$        & $3.085\pm0.033$                          &   $3.076^{+0.035}_{-0.036}$           &   $3.115\pm0.038$           & $3.078\pm0.033$   &  $3.081\pm0.033$ &   $3.084\pm0.036$   &  $3.115\pm0.042$\\

\hline
$\Omega_{\rm M}$  &     $0.2796\pm0.0097$         &    $0.2804^{+0.0103}_{-0.0105}$          &    $0.2829^{+0.0100}_{-0.0102}$          &  $0.2767^{+0.0108}_{-0.0107}$  &  $0.2815^{+0.0130}_{-0.0128}$ &  $0.2821^{+0.0106}_{-0.0107}$     & $0.2798^{+0.0146}_{-0.0144}$\\
$100h$  &     $69.72^{+0.90}_{-0.91}$         &     $69.40^{+1.03}_{-0.99}$         &         $74.14^{+3.17}_{-3.14}$     &  $70.46^{+1.44}_{-1.38}$  & $70.56\pm1.43$  &   $69.33^{+1.03}_{-1.01}$   & $73.78^{+3.16}_{-3.17}$\\
$\sigma_8$  &   $0.816\pm0.021$           &      $0.772^{+0.036}_{-0.034}$         &       $0.807\pm0.044$      &  $0.786\pm0.046$  &  $0.792\pm0.053$ &   $0.772\pm0.039$    &  $0.796^{+0.063}_{-0.064}$\\
Age [Gyrs]  &      $13.79\pm0.10$        &    $13.84\pm0.12$           &    $12.95^{+0.53}_{-0.54}$        & $13.86\pm0.12$   & $13.84\pm0.12$  &    $13.83\pm0.11$     & $13.28\pm0.59$ \\


\hline\hline
\end{tabular}
\end{center}
\caption{The mean and the 68 percent CL errors for the cosmological parameters measured assuming various models. For the parameters with a one-tailed distribution, such as the neutrino mass and $r$, the 95 percent CL upper limit is presented instead.}
\label{tab:mnu_models}
\end{table*}%

\begin{table*}

\begin{center}
\begin{tabular}{c|c|c|c|cc}

\hline\hline
                                    & \multicolumn{2}{c}{Minimal model}                         &    \multicolumn{3}{c}{All float} \\ 
                                    &  Parametrisation  & Constraint [CMASS $P(k)$]                       &    Parametrisation  & Constraint [CMASS $P(k)$]  & Constraint [CMASS BAO] \\
                                                                        
                                    \hline
                                    
$\Sigma m_{\nu}$ [eV]      &  $\Lambda m_{\nu}$CDM  &  $<0.340$      & All float   &$<0.821$  & $<1.143$\\
$N_{\rm eff}$                     &  $\Lambda N_{\rm eff} m_{\nu}$CDM &  $4.308\pm0.794$       & All float  &$4.032^{+0.870}_{-0.894}$ & $4.324\pm0.881$   \\
$w_0$                             &  $w_0 w_a$CDM &  $-1.041\pm0.143$      & All float & $-0.964^{+0.173}_{-0.168}$ &  $-0.899^{+0.162}_{-0.167}$  \\
$w_a$                         &  $w_0 w_a$CDM    &          $-0.111\pm0.708$ & All float  & $-0.731^{+0.982}_{-1.053}$ &  $-1.455^{+0.997}_{-1.010}$\\
$100\Omega_K$          &  $o$CDM &   $-0.264^{+0.466}_{-0.461}$   & All float &$-0.411\pm1.02$  & $-0.631^{+0.902}_{-0.889}$\\
$r$                   &  $\Lambda r$CDM & $<0.198$       & All float & $<0.440$   & $<0.405$\\
$\alpha_s$       &  $\Lambda\alpha_s$CDM&       $-0.017^{+0.018}_{-0.017}$ & All float & $-0.032\pm0.030$  & $-0.023\pm0.030$\\    \hline                                
                                    
$100\Omega_b h^2$   &  $\Lambda$CDM &      $2.265\pm0.050$                 & All float & $2.268^{+0.080}_{-0.081}$     &  $2.254^{+0.082}_{-0.083}$ \\
$100\Omega_c h^2$   &  $\Lambda$CDM &      $11.32^{+0.243}_{-0.245}$         & All float &  $12.974^{+1.304}_{-1.329}$  & $13.671^{+1.587}_{-1.575}$\\
$\Theta_s$                  &  $\Lambda$CDM &     $1.0397^{+0.0024}_{-0.0025}$    & All float  &     $1.0364\pm0.0039$ & $1.0351\pm0.0038$\\
$\tau$                          &  $\Lambda$CDM &      $0.0870^{+0.0063}_{-0.0070}$ & All float   &  $0.0927^{+0.0069}_{-0.0082}$ &  $0.0922^{+0.0069}_{-0.0079}$\\
$n_s$                          &  $\Lambda$CDM &       $0.967\pm0.012$                     & All float   & $0.965^{+0.043}_{-0.042}$ & $0.971^{+0.042}_{-0.041}$\\
ln$[10^{10} A_s]$       &  $\Lambda$CDM & $3.085\pm0.033$                          & All float & $3.115\pm0.042$  & $3.121\pm0.045$\\

\hline
$\Omega_{\rm M}$  &  $\Lambda$CDM &     $0.2796\pm0.0097$      & All float         &  $0.2798^{+0.0146}_{-0.0144}$ & $0.2861^{+0.0152}_{-0.0153}$\\
$100h$  &  $\Lambda$CDM &     $69.72^{+0.90}_{-0.91}$           & All float    &  $73.78^{+3.16}_{-3.17}$  & $74.56^{+3.44}_{-3.34}$\\
Age [Gyrs]  &  $\Lambda$CDM &      $13.79\pm0.10$           & All float     &  $13.28\pm0.59$ & $13.20^{+0.64}_{-0.66}$\\

\hline\hline
\end{tabular}
\end{center}
\caption{The constraint of the cosmological parameters in the optimistic (vanilla parameters with a minimal extension to include the parameter concerned) and the conservative case where all parameters are allowed to float. For the conservative case, we show the result using the CMASS $P(k)$ and CMASS BAO respectively.}
\label{tab:summery}
\end{table*}%

In this section, we present the measurements of other cosmological parameters. For each parameter $x$, we first present the constraint in the $x$CDM model\footnote{If $x$ is a member of the set of the vanilla parameters presented in Table~\ref{tab:param}, we present the constraint in $\Lambda$CDM.}. We then provide the constraint in either the $x m_{\nu}$CDM, or the `All float' models, according to the degeneracy between $x$ and the neutrino mass. In this way, we can see the impact of neutrino mass, and other cosmological parameters on the measurement of parameter $x$.

\subsubsection{$\Omega_{\rm M}$ and $H_0$}

\begin{figure}
\begin{center}
\includegraphics[scale=0.3]{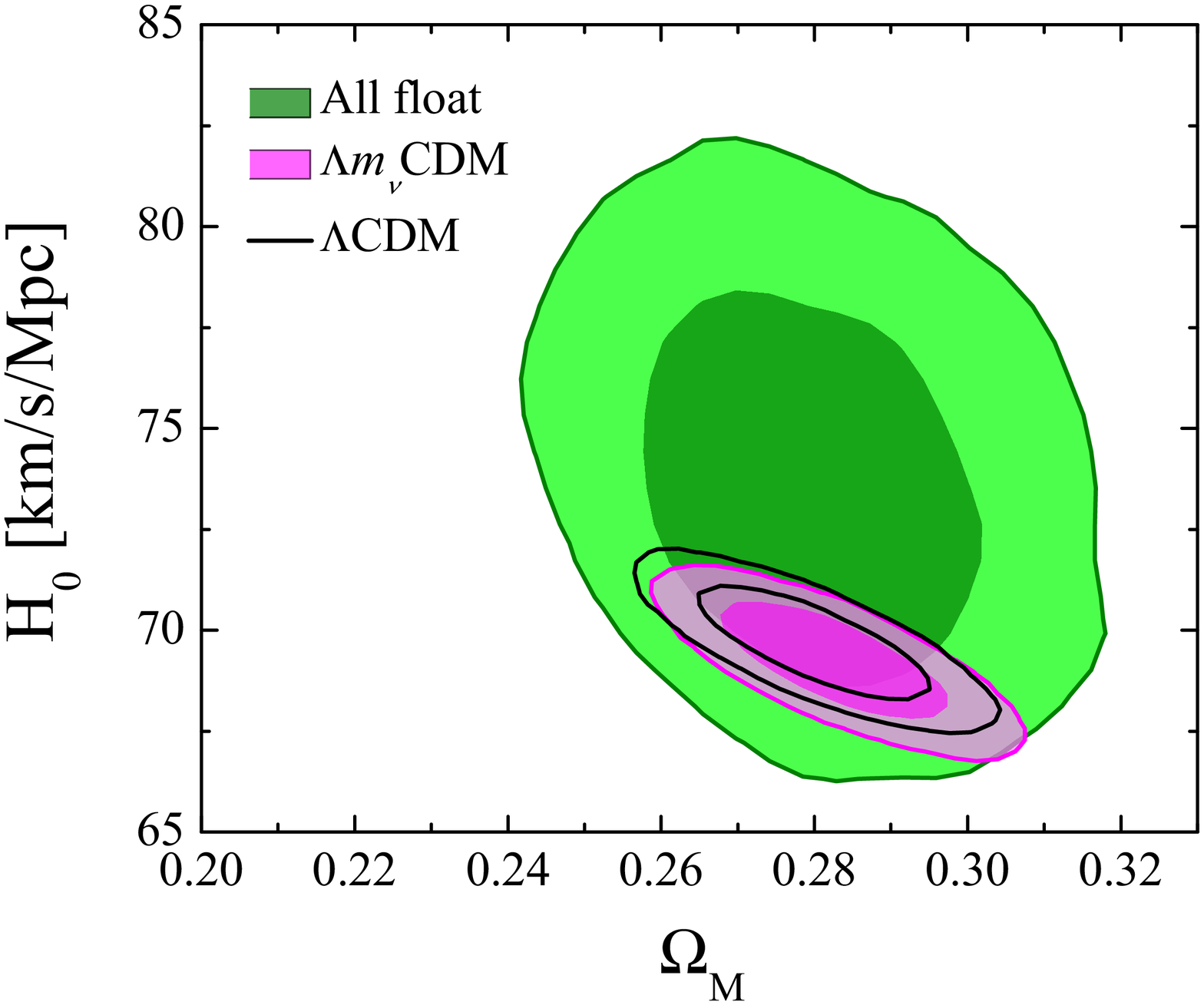}
\end{center}
\caption{The 68 and 95 percent CL contour plots for $\Omega_{\rm M}$ and $H_0$ obtained from the joint dataset for $\Lambda$ CDM (black solid),  $\Lambda m_{\nu}$ CDM (magenta filled) and `All float' models (green filled).}\label{fig:omm_H0}
\end{figure}

In $\Lambda$CDM, $\Omega_{\rm M}$ and $H_0$ are well determined, namely, 
\begin{equation} \Omega_{\rm M}=0.2796\pm0.0097,~~H_0=69.72^{+0.90}_{-0.91}~{\rm km/s/Mpc}\end{equation}
If the neutrino mass is marginalised over, the constraint becomes,
\begin{equation} \Omega_{\rm M}=0.2804^{+0.0103}_{-0.0105},~~H_0=69.40^{+1.03}_{-0.99}~{\rm km/s/Mpc}\end{equation}
If all parameters are marginalised over, the constraint is diluted to,
\begin{equation} \Omega_{\rm M}=0.2798^{+0.0132}_{-0.0136},~~H_0=73.78^{+3.16}_{-3.17}~{\rm km/s/Mpc}\end{equation}
From these numbers and also Fig.~\ref{fig:omm_H0}, we can see that marginalising over neutrino mass can only dilute the constraint on these background parameters by roughly 10 percent, but if other parameters including $w_0,w_a,\Omega_K$ vary, the degradation is significant.   

\subsubsection{Curvature $\Omega_K$}

\begin{figure}
\begin{center}
\includegraphics[scale=0.2]{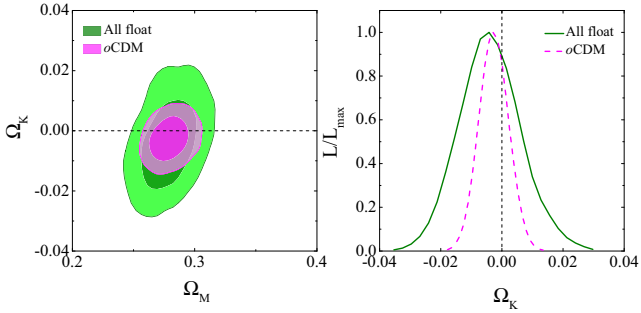}
\end{center}
\caption{Left: green contours on the back layer: the 68 and 95 percent CL constraints for $\Omega_K$ and $\Omega_{\rm M}$ in the `All float' cosmology. Transparent magenta contours on the top layer: the same for the $o$CDM cosmology; Right: the corresponding $1D$ posterior distribution of $\Omega_K$. The vertical black dashed line shows $\Omega_K=0$ to guide eyes.}\label{fig:omk}
\end{figure}

The $(\Omega_K,\Omega_{\rm M})$ contour and the posterior distribution of $\Omega_K$ are shown in Fig.~\ref{fig:omk}. In the `All float' model, the $\Omega_K$ constraint is weakened by more than a factor of $2$, namely, $\Omega_K=-0.00264^{+0.00466}_{-0.00461}$ ($o$CDM), and $\Omega_K=-0.00411\pm{0.01029}$ (All float). But in both cases, $\Omega_K$ is very consistent with zero. 

\subsubsection{The constant $w$}

Due to the degeneracy between $w$ and the summed neutrino mass shown in Fig.~\ref{fig:w_mnu}, the constraint of $w$ is slightly relaxed when the summed neutrino mass is marginalised over, namely, $w=-1.068\pm0.072$ ($w$CDM); $w=-1.081\pm0.075$ ($w m_{\nu}$CDM). If all the parameters are marginalised over, the constraint is further weakened to $w=-1.067\pm0.121$.

\begin{figure}
\begin{center}
\includegraphics[scale=0.25]{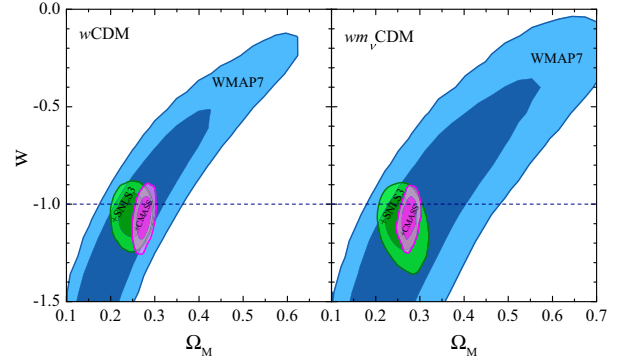}
\end{center}
\caption{The 68 and 95 percent CL contour plots for $w$ and $\Omega_{\rm M}$ obtained from the joint dataset including CMB (blue), CMB+SN (green), CMB+SN+CMASS power spectrum (magenta). The neutrino mass is fixed to zero and allowed to vary in the left and right panels respectively.}\label{fig:w_omm}
\end{figure}

\subsubsection{$w_0$ and $w_a$}

\begin{figure}
\begin{center}
\includegraphics[scale=0.3]{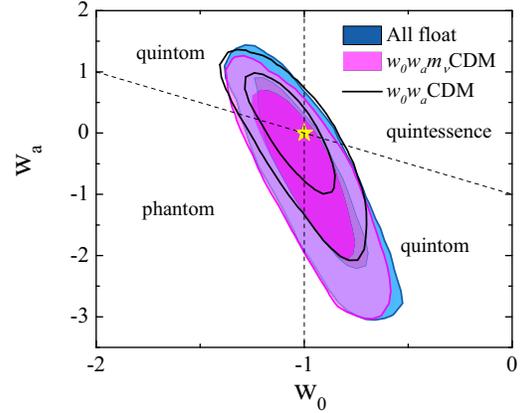}
\end{center}
\caption{The 68 and 95 percent CL contour plots for $w_0$ and $w_a$ obtained from the joint dataset in three cosmologies: $w_0 w_a$CDM (black curves), $w_0 w_a m_{\nu}$CDM (magenta) and `All float' (blue). The black dashed lines subdivide the parameter space according to the theoretical dark energy model predictions, and the yellow star illustrates the $\Lambda$CDM model. }\label{fig:w0wa}
\end{figure}

Similarly, the CPL parameters $w_0$ and $w_a$ are less constrained when the other cosmological parameters, including the neutrino mass, are simultaneously fitted: 
\begin{eqnarray} 
  &&w_0=-1.041\pm0.143,~w_a=-0.111\pm0.708~(w_0w_a{\rm CDM}), \nonumber \\     
  &&w_0=-0.984\pm0.157,~w_a=-0.704\pm0.901~(w_0w_a m_{\nu}{\rm CDM}), \nonumber \\   
  &&w_0=-0.964\pm0.168,~w_a=-0.731\pm0.970~({\rm All~float}). \nonumber 
\end{eqnarray} 
The contour plot for $w_0,~w_a$ is shown in Fig.~\ref{fig:w0wa} for three cosmologies. As we can see, marginalising over the neutrino mass can significantly dilute the measurement of $w_0$ and $w_a$, and also shift the contour towards a more negative $w_a$. This is understandable -- when the massive neutrino, whose equation-of-state is non-negative, is assumed to exist in the universe, a dark energy component with a more negative equation-of-state (assumed constant) is needed to compensate. This is why we see an anti-correlation between $w$ and the neutrino mass in Fig \ref{fig:w_mnu}. If the dynamics of dark energy, represented by a nonzero $w_a$ for example, is allowed, the neutrino mass anti-correlates with the effective equation-of-state $w_{\rm eff}$ of dark energy (see Eq(57) in \citealt{DEP}). As a result, either $w_0$ or $w_a$ needs to be more negative if $\Sigma m_{\nu}\ne0$.

\subsubsection{$N_{\rm eff}$} 

\begin{figure}
\begin{center}
\includegraphics[scale=0.2]{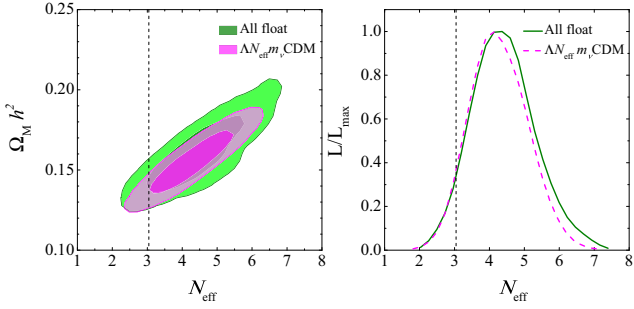}
\end{center}
\caption{Left panel: the 68 and 95 percent CL contour plots for $N_{\rm eff}$ and $\Omega_{\rm M} h^2$ obtained from the joint dataset for $\Lambda N_{\rm eff} m_{\nu}$CDM (magenta) and `All float' models (green); Right panel: the corresponding $1D$ posterior distribution of $N_{\rm eff}$.}\label{fig:Neff}
\end{figure}

The number of neutrino species is measured to be $4.308\pm0.794$ and $N_{\rm eff}=4.412^{+0.865}_{-0.876}$, in the $\Lambda N_{\rm eff} m_{\nu}$CDM and `All float' models respectively. In both cases, the standard value $N_{\rm eff}=3.046$ is only consistent with this at the $1.6\sigma$ level. 

Fig.~\ref{fig:Neff} shows the contour plot of $N_{\rm eff}$ and $\Omega_{\rm M} h^2$, and the posterior distribution of $N_{\rm eff}$ for both $\Lambda N_{\rm eff} m_{\nu}$CDM and `All float' models. As we see, $N_{\rm eff}$ strongly correlates with $\Omega_{\rm M} h^2$ as expected since $N_{\rm eff}$ is directly related to $\Omega_{\rm M} h^2$ (see Eq.~(53) in \citealt{Komatsu11}).

\subsubsection{Tensor mode and the inflation models}

\begin{figure}
\begin{center}
\includegraphics[scale=0.295]{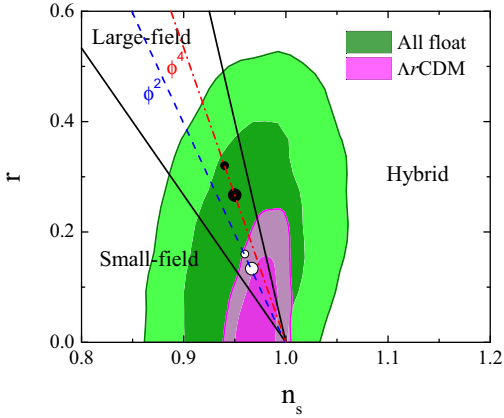}
\end{center}
\caption{68 and 95 percent CL contours for $(n_s,r)$ in the $\Lambda r$CDM (magenta) and the `All float' models (green). The solid black lines subdivide the parameter space to show different inflation model predictions. The blue dashed and red dash-dotted lines illustrate the models of $\phi^2$ and $\phi^4$ models respectively. The small and large dots on each line show the model prediction for the e-fold $N=50$ and $60$ respectively. See text for more details.}\label{fig:ns_r}
\end{figure}

\begin{figure}
\begin{center}
\includegraphics[scale=0.2]{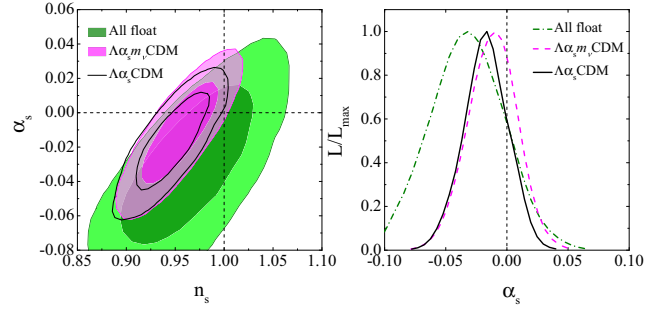}
\end{center}
\caption{Left panel: 68 and 95 percent CL contours for $n_s$ and $\alpha_s$ obtained from the joint dataset for $\Lambda\alpha_s$ CDM (black solid),  $\Lambda\alpha_s m_{\nu}$ CDM (magenta filled) and `All float' models (green filled); Right panel: the corresponding $1D$ posterior distribution of $\alpha_s$.}\label{fig:alphas}
\end{figure}

\begin{figure*}
\begin{center}
\includegraphics[scale=0.4]{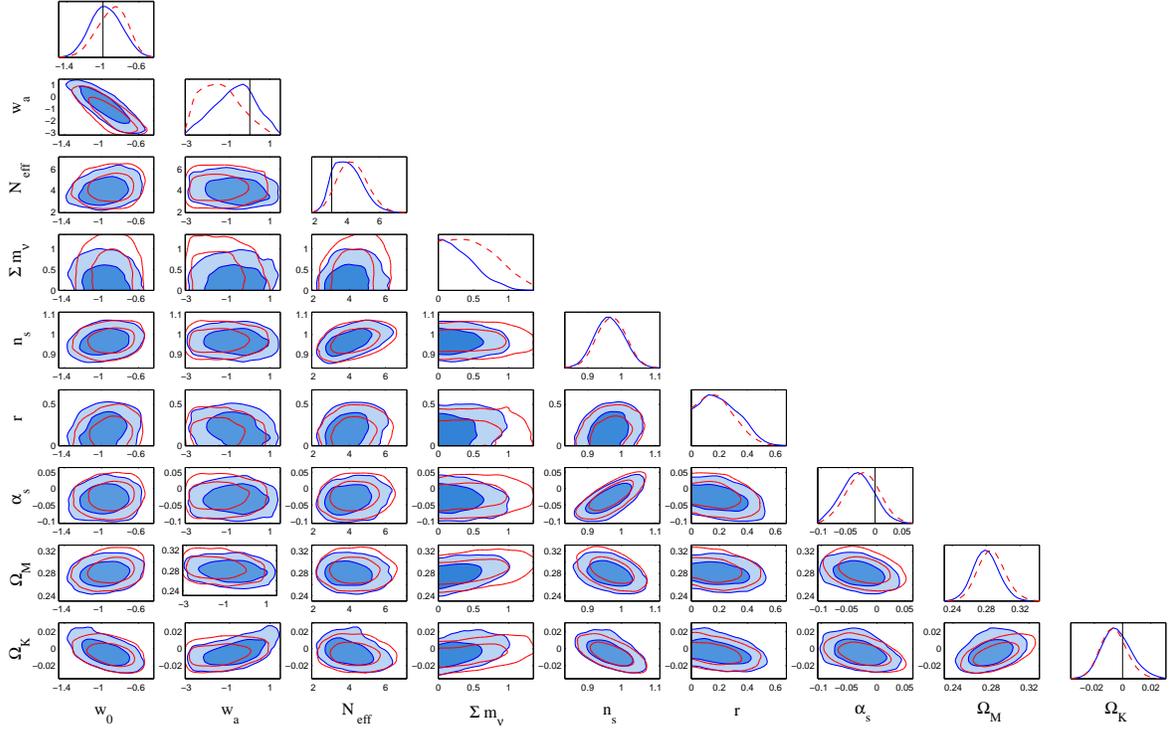}
\end{center}
\caption{The $1D$ posterior distribution of the cosmological parameters listed in Table~\ref{tab:param} and the $2D$ 68 and 95 percent CL contours within them. The shaded contours and blue solid curves show the results calculated using the CMASS power spectrum, while the unfilled contours and the red dashed curves show the result using the CMASS BAO distance measurements. The CMB, SN and other BAO datasets are combined for both cases. The black vertical lines show the standard values for some of the parameters.}\label{fig:tri}
\end{figure*}

Various inflation models can be classified on the $(n_s,r)$ plane. For example, the Small-field and Large-field models predict $r<\frac{8}{3}(1-n_s),~n_s\leqslant1$ and $\frac{8}{3}(1-n_s)<r<8(1-n_s),~n_s\leqslant1$, respectively, while Hybrid inflation predicts $r>\frac{8}{3}(1-n_s)$ and $n_s$ free \citep{Peiris03,Tegmark04}. To test various inflation models, we show the $(n_s,r)$ contour plot in Fig.~\ref{fig:ns_r}. The constraint is
\begin{eqnarray} 
  &&n_s=0.982\pm0.012,~~r<0.198~~(\Lambda r{\rm CDM}), \nonumber \\ 
  &&n_s=0.965^{+0.043}_{-0.042},~~~~~~r<0.440~~({\rm All~float}).
\end{eqnarray} 
We also show the prediction of two Large-field inflation models, 
\begin{itemize}
  \item $V\propto\phi^2:~r=\frac{8}{N},~n_s=1-\frac{2}{N}$,
  \item $V\propto\phi^4:~r=\frac{16}{N},~n_s=1-\frac{3}{N}$,
\end{itemize}
where $\phi,V$ denote the inflaton and the potential of the inflaton respective, and $N$ stands for the e-fold number.  We show the model predictions for both the $\phi^2$ and $\phi^4$ cases for $N=50,60$. As we can see, in the $\Lambda r$CDM cosmology, the $\phi^4$ model is significantly disfavoured, but the $\phi^2$ model is still allowed. However, both models are consistent with observations for the case where all the parameters are allowed to vary. The scale-invariant spectrum ($n_s=1$) is outside the 68 percent CL region of the resulting contour, but is within the 95 percent CL region in both cases.    
 
\subsubsection{Running of the primordial power spectrum}

\begin{table*}

\begin{center}
\begin{tabular}{c|c|c}

\hline\hline
$\Sigma m_{\nu}$ [eV]& Reference &  Galaxy data used\\ \hline
$<0.80~(0.81)$ & \citet{Saito:2009fk} &  $3D$ power spectrum of SDSS-II LRG\\
$<0.30~(0.51)$ & \citet{Reid10b}       &    maxBCG \\
$<0.28~(0.47)$ &\citet{Thomas10} & MegaZ SDSS DR7 \\
$<0.51$ & \citet{Sanchez:2012sg} &  SDSS-III CMASS two-point correlation function combined with BAO of other surveys\\
$<0.26~(0.36)$ & \citet{dePutter:2012sh} &  SDSS-III DR8 LRG angular power spectrum \citep{DR8, Ho12} \\
$<0.29~(0.41)$ & \citet{Xia12} &   Angular power spectrum of CFHTLS galaxy counts  \\
$<0.48~(0.63)$ & \citet{2012arXiv1210.2136W} &   Weak lensing measurement of the CFHTLS-T0003 sample \\
$<0.32$ & \citet{2012arXiv1210.2130P} &   Angular power spectrum of WiggleZ galaxy counts  \\
$<0.340~(0.821)$ & This work &  SDSS-III CMASS $3D$ power spectrum combined with BAO of other surveys\\
\hline\hline

\end{tabular}
\end{center}
\caption{Recently published neutrino mass constraints found in the literature. For the neutrino mass constraints, the number in the parentheses shows the measurement obtained under more conservative assumptions, where available.}
\label{tab:refs}
\end{table*}%
The running of the primordial power spectrum, $\alpha_s$, is constrained to be $\alpha_s=-0.017^{+0.018}_{-0.017}$ in the $\Lambda\alpha_s$CDM cosmology. Since both $\alpha_s$ and the neutrino mass can change the shape of $P(k)$ \citep{Feng06}, we redo the analysis with the neutrino mass also allowed to vary. The resultant measurement is slightly diluted, namely, $\alpha_s=-0.012\pm0.019$. When all the parameters vary, the constraint is further weakened to be $\alpha_s=-0.032\pm0.030$.  The contour plot for $n_s$ and $\alpha_s$ and the $1D$ posterior distribution of $\alpha_s$ are presented in Fig.~\ref{fig:alphas}. As shown, $\alpha_s$ is consistent with zero in all of the cases we consider.

\subsubsection{Measuring all of the parameters simultaneously}

Measurements for all of the cosmological parameters are listed in Table \ref{tab:summery}. For each parameter $x$, we consider both the cases where the varied parameters are the vanilla parameters with a minimal extension to include the parameter concerned, and the measurement in the `All float' model, when all parameters are simultaneously fitted. For the 'All float' model, we compare the result using the CMASS $P(k)$ with that using only the CMASS BAO information. Comparison of these measurements shows the additional information coming from the power spectrum, even when allowing for nuisance parameters for the galaxy bias modelling using Eq.~(\ref{eq:nonlinearPkhalo}). 

The $1D$ posterior distribution of the cosmological parameters and the $2D$ contours for pairs of these parameters are shown in Fig.~\ref{fig:tri}. The constraints using BAO only are generally weaker showing that the shape information in $P(k)$ isn't significantly degraded even if a sophisticated galaxy bias modelling with several nuisance parameters is used. 

\section{Conclusions}

We have measured the summed neutrino mass using the power spectrum of the SDSS-III CMASS sample. We also investigate how the prior knowledge of the neutrino mass affects the measurement of other cosmological parameters. We do this by studying both the minimal model, where we have a flat, $\Lambda$CDM background with the `extra' parameter considered, and also study the bounds on cosmological parameters in a more general case, in which all parameters are allowed to vary simultaneously.  

For the neutrino mass constraint, we discuss several factors which might affect the final result, including the choice of SN data, the treatment of the galaxy modelling, and the cutoff scale of the power spectrum for the analysis. Our main results can be summarised as follows,    

\begin{enumerate}

\item {\it How heavy are the neutrinos?} \\

When we assume a $\Lambda m_{\nu}$CDM cosmology, we find that $\Sigma m_{\nu}<0.340$\,eV (95 percent CL) using a joint analysis of data from WMAP7, SNLS3, the CMASS power spectrum, and the BAO measurement of WiggleZ, 6dF and SDSS-II. In the most conservative case, where all the cosmological parameters in Table~\ref{tab:param} are allowed to vary simultaneously, the constraint is weakened to $\Sigma m_{\nu}<0.821$\,eV (95 percent CL). 

There are several recently published neutrino mass constraints listed in Table \ref{tab:refs}. The more conservative constraints, which was derived using a smaller $k_{\rm max}$, or assuming a more general cosmology and so on, are also listed in the parentheses.  The difference among these measurements is due to different datasets used and different galaxy modelling adopted.  The conservative result of this work, $\Sigma m_{\nu}<0.821$\,eV is significantly weaker showing how freedom in other parameters degrades $\Sigma m_{\nu} $ measurements, while the other measurements were done in the $\Lambda m_{\nu}$CDM cosmology. \\

\item {\it How many neutrino species?} \\

The number of neutrino species is measured to be $4.308\pm0.794$ in the $\Lambda N_{\rm eff} m_{\nu}$CDM. In this case, the standard value $N_{\rm eff}=3.046$ is only allowed at the $1.6\sigma$ level. In the `All float' model, the constraint is degraded to $N_{\rm eff}=4.412^{+0.865}_{-0.876}$, but the standard value is still only accepted at the $1.57\sigma$ level. 

Interestingly, recently several groups have also measured $N_{\rm eff}>3.046$, including the Atacama Cosmology Telescope (ACT) \citep{ACT} and the South Pole Telescope (SPT) \citep{SPT} teams. This allows the possibility of the existence of one additional neutrino species, \ie, the sterile neutrino (\eg, \citealt{Ciuffoli12}).\\ 

\item {\it Is $w=-1$?}\\

If one fits a constant $w$ to the joint dataset, then $w=-1.068\pm0.072$ ($w$CDM) and $w=-1.067\pm0.121$ (All float). For the CPL parametrisation, which allows $w$ to vary linearly in the scale factor $a$, $w=-1$ is still well within the 95 percent CL limit. 

The dynamics of $w$ can be further probed by using a
more general parameterisation, or a non-parametric approach to reconstruct $w(a)$. In fact, \citet{Zhao12} recently performed a non-parametric reconstruction of $w(a)$ using a similar dataset and they found a signal for the dynamics of dark energy at $>2\sigma$ level. In principle, this is not necessarily inconsistent with our result here simply because the CPL parametrisation we used in this work is not general enough to cover the behaviour of $w(a)$ that \citet{Zhao12} found. Actually, the CPL parametrisation provides a much worse fit to the data (the $\chi^2$ is only reduced by 0.5 compared to the $\Lambda$CDM model) than the \citet{Zhao12} result, which reduces the $\chi^2$ by $6$ with 3 effective number of degrees of freedom. \\

\item{\it Is the universe close to flat?}\\

The answer is yes based on this analysis. The conclusion that $\Omega_K=0$ is robust -- the measured $\Omega_K$ is very consistent with zero in all the cases we considered. \\

\item{\it Is there a running in the primordial power spectrum?} \\

The answer is no. We confirm that $\alpha_s$ is consistent with zero at the 95 percent CL in all the cases we studied.  \\

\item{\it How much tensor perturbation is allowed?} \\

The tensor-to-scalar ratio is tightened down to $r<0.198$ and $r<0.440$ in the minimal cosmological model and the 'Full fit' model respectively. This rules out the $\phi^2$ inflation model with the e-fold number $N<60$ to more than $3\sigma$. However, the $\phi^4$ model with $N\gtrsim50$ is still allowed. 

\end{enumerate}

All the above issues can be further investigated using the accumulating BOSS data in the next few years.   

\section*{Acknowledgements}

We thank Uros Seljak, Kyle Story, Licia Verde and Yvonne Wong for discussions.

Funding for SDSS-III has been provided by the Alfred P. Sloan Foundation, the Participating Institutions, the National Science Foundation, and the U.S. Department of Energy Office of Science. The SDSS-III web site is \url{http://www.sdss3.org/}.

MV is supported by the ERC-StG cosmoIGM.

SDSS-III is managed by the Astrophysical Research Consortium for the Participating Institutions of the SDSS-III Collaboration including the University of Arizona, the Brazilian Participation Group, Brookhaven National Laboratory, University of Cambridge, Carnegie Mellon University, University of Florida, the French Participation Group, the German Participation Group, Harvard University, the Instituto de Astrofisica de Canarias, the Michigan State/Notre Dame/JINA Participation Group, Johns Hopkins University, Lawrence Berkeley National Laboratory, Max Planck Institute for Astrophysics, Max Planck Institute for Extraterrestrial Physics, New Mexico State University, New York University, Ohio State University, Pennsylvania State University, University of Portsmouth, Princeton University, the Spanish Participation Group, University of Tokyo, University of Utah, Vanderbilt University, University of Virginia, University of Washington, and Yale University.

\bibliographystyle{mn2e}
\bibliography{draftv2}

\label{lastpage}

\end{document}